\documentclass[aps,prb,twocolumn,10pt,superscriptaddress,showpacs]{revtex4-2}
\usepackage{amsmath,amssymb,graphics,epsfig,epstopdf,color,verbatim,ulem,braket,tabularx}
\usepackage[colorlinks,linkcolor=blue,citecolor=blue,urlcolor=blue]{hyperref}

\begin{document}

\title{Nature of the mixed-parity pairing of attractive fermions with spin-orbit coupling in optical lattice}

\author{Yu-Feng Song}
\affiliation{Hefei National Laboratory for Physical Sciences at Microscale and Department of Modern Physics, University of Science and Technology of China, Hefei, Anhui 230026, China}
\affiliation{Institute of Modern Physics, Northwest University, Xi'an 710127, China}

\author{Youjin Deng}
\email{yjdeng@ustc.edu.cn}
\affiliation{Hefei National Laboratory for Physical Sciences at Microscale and Department of Modern Physics, University of Science and Technology of China, Hefei, Anhui 230026, China}
\affiliation{Hefei National Laboratory, University of Science and Technology of China, Hefei 230088, China}

\author{Yuan-Yao He}
\email{heyuanyao@nwu.edu.cn}
\affiliation{Institute of Modern Physics, Northwest University, Xi'an 710127, China}
\affiliation{Shaanxi Key Laboratory for Theoretical Physics Frontiers, Xi'an 710127, China}
% \affiliation{Peng Huanwu Center for Fundamental Theory, Xian 710127, China}
\affiliation{Hefei National Laboratory, University of Science and Technology of China, Hefei 230088, China}

\begin{abstract}
The admixture of spin-singlet and spin-triplet pairing states in superconductors can be typically induced by breaking spatial inversion symmetry. Employing the {\it numerically exact} auxiliary-field Quantum Monte Carlo method, we study such mixed-parity pairing phenomena of attractive fermions with Rashba spin-orbit coupling (SOC) in two-dimensional optical lattice at finite temperature. We systematically explore the evolution of the essential pairing structure in both singlet and triplet channels versus the temperature, fermion filling, SOC and interaction strengths, via computing the finite-size results of condensate fraction and pair wave function. Our numerical results reveal that the singlet channel dominates in the fermion pairing and the triplet pairing has relatively small contribution to the superfluidity for physically relevant parameters. In contrast to the singlet channel mainly consisted of the on-site Cooper pairs, the triplet pairing has plentiful patterns in real space with the largest contributions from several nearest neighbors. As the SOC strengh increases, the pairing correlation is firstly enhanced and then suppressed for triplet pairing while it's simply weakened in singlet channel. We have also obtained the Berezinskii-Kosterlitz-Thouless transition temperatures through the finite-size analysis of condensate fraction. Our results can serve as quantitative guide for future optical lattice experiments as well as accurate benchmarks for theories and other numerical methods. 
\end{abstract}

\date{\today}
\maketitle

\section{Introduction}
\label{sec:intro}

The fermion paring and corresponding superconductivity and superfluidity~\cite{Ginzburg2004} are of great interest in condensed matter physics. The fundamental ingredient is the Cooper pair consisting of two spin-$1/2$ electrons~\cite{Cooper1956}. Given the spatial inversion symmetry, the pair wave function can be decoupled into orbital and spin channels resulting in two states of Cooper pairs, even parity with spin-singlet and odd parity with spin-triplet~\cite{Anderson1984}. Majority of known superconductors (SCs) fall into the spin-singlet case, such as simple metals~\cite{Bardeen1957} and high-$T_c$ cuprates~\cite{Tsuei2000,*Lee2006,*Armitage2010}. Nevertheless, the triplet paring has been observed or suggested to exist in far fewer realistic systems, e.g., superfluid $^3$He~\cite{Lee1997}, $\text{UPt}_3$ and $\text{Sr}_2\text{RuO}_4$~\cite{Mackenzie2003}. Without inversion symmetry, the parity conservation is broken and thus the mixing of singlet and triplet paring states can emerge~\cite{Rashba2001,Frigeri2004,Yip2014,Smidman2017}. Such mixed-parity pairing state has been experimentally verified in various three-dimensional (3D) noncentrosymmetric SCs~\cite{Bauer2004,Yuan2006,Nishiyama2007,Xu2020,Ishihara2021,Cameron2022}, which induces intensive interests due to many exotic properties~\cite{Yip2014} including fertile superconducting gap structures~\cite{Yuan2006,Bonalde2005}, anisotropic magnetic response~\cite{Agterberg2004,Samokhin2005} and topological superconductivity~\cite{Gao2022}.

The appearance of the mixed-parity pairing in noncentrosymmetric systems can be attributed to the arise of the antisymmetric spin-orbit coupling (SOC)~\cite{Bychkov1984}, which has become one of the key elements for condensed matter physics~\cite{Manchon2015}. For correlated fermion systems, SOC acts as another dimension and induces many exotic states of matter, including spintronics~\cite{Atsufumi2020}, topological phases~\cite{Hasan2010,*Qi2011} and unusual superconductivity~\cite{Smidman2017}. Specifically, it typically breaks the spatial inversion symmetry and mixes the spin species, and thus renders the coexistence of spin-singlet even-parity and spin-triplet odd-parity pairing. Moreover, it was shown~\cite{Nishiyama2007} that tunning the SOC strength can even change the dominant component of the mixed-pairity pairing from the singlet in $\text{Li}_2\text{Pd}_3\text{B}$ to the triplet in $\text{Li}_2\text{Pt}_3\text{B}$, as replacing the Pd atom by Pt atom. To date, most of the study for the SOC induced singlet-triplet mixed pairing phenomena concentrates on the 3D systems including the noncentrosymmetric SCs~\cite{Smidman2017} and interacting Fermi gas~\cite{Hu2011,Zhou2012,Powell2022}.

In physically more relevant two-dimensional (2D) systems, the interplay between the reduced dimensionality and enhanced quantum fluctuations can induce fascinating and unique quantum phenomena~\cite{Arovas2021,Bloch2008,*Giorgini2008,Castro2009,*Kotov2012}. A typical representative is the Berezinskii-Kosterlitz-Thouless (BKT) transition~\cite{Berezinsky1972,Kosterlitz1973,Jorge2013,Ryzhov2017} of superconductivity and superfluidity. Similar to the 3D analog, inclusion of SOC to 2D attractive fermions can also induce mixed-pairity pairing~\cite{Rashba2001}, which has been relatively much less studied. Experimentally, the recently elegant realization of synthetic SOC for fermions~\cite{Cheuk2012,Wang2012} with ultracold atoms, especially in 2D optical lattice~\cite{Huang2016,Meng2016,Sun2018}, substantively paves the way for exploring novel quantum phenomena related with SOC. Thus, a systematically theoretical study with high precision on the mixed-pairity pairing in 2D is highly demanded to shed light on problems closely related to ultracold atom experiments. For example, finding the best condition to observe the spin-triplet pairing in optical lattice, in comparison to the achieved singlet pairing~\cite{Hartke2023}, should be a useful guide for experiments. 

To date, most theoretical work on 2D attractive fermions with SOC falls into the Fermi gas and approximate theories~\cite{Anna2011,*Anna2012,Sun2013,Devreese2014,Devreese2015,Lee2017}. Interesting results such as singlet and triplet contributions to the condensation~\cite{Anna2011,*Anna2012} are presented in these studies, but still need careful verifications from unbiased approaches. Nevertheless, numerically exact calculations for such systems are rare. Auxiliary-field Quantum Monte Carlo (AFQMC) simulations have been performed for the ground state of 2D Fermi gas~\cite{Shi2016}, and for the lattice system at finite temperatures~\cite{Tang2014} as well as its ground state~\cite{Rosenberg2017}. The authors in Ref.~\cite{Tang2014} focused on the properties of BKT transition temperatures and anisotropic spin susceptibility without touching the pairing structure, which were limited to 12$\times$12 finite lattices. The pairing structure were discussed in Ref.~\cite{Rosenberg2017} only for the half-filling case, for which the BKT transition disappears and thus it was of less interest to experiments. 

In this paper, we study the mixed-parity pairing of attractive fermions with Rashba SOC in 2D optical lattice, applying finite-temperature AFQMC algorithm~\cite{Blankenbecler1981,Hirsch1983,White1989,Yuanyao2019}. We mainly concentrate on the condensate fraction and pair wave functions to reveal the pairing structures of both singlet and triplet channels for physically relevant regimes of the temperature, fermion filling, SOC and interaction strengths. We also present the determination of BKT transition temperature from the finite-size analysis of condensate fraction results. The rest of the paper is organized as follows. In Sec.~\ref{sec:modelmethod}, we introduce the lattice model that we use to describe the 2D attractive fermions with Rashba SOC in optical lattice, and the AFQMC method. In Sec.~\ref{sec:QMCresults}, we present our numerical results, including the pairing structures, the pairing correlations and calculations of the BKT transition temperature. Finally, Sec.~\ref{sec:Summary} summarizes this work, and discusses its connections with the optical lattice experiments.

\section{Model and method}
\label{sec:modelmethod}

We describe the 2D attractive fermions with Rashba SOC in optical lattice using the following square lattice model Hamiltonian~\cite{Tang2014,Rosenberg2017} as
\begin{equation}\begin{aligned}
\label{eq:2DHamlt}
\hat{H}=&\sum_{\mathbf{k} \sigma} \varepsilon_{\mathbf{k}} c_{\mathbf{k}\sigma}^{\dag}c_{\mathbf{k}\sigma}+\sum_{\mathbf{k}} (\mathcal{L}_{\mathbf{k}}c_{\mathbf{k}\downarrow}^{\dag}c_{\mathbf{k}\uparrow}+\mathrm{H.c.}) \\
&\hspace{0.3cm}+U\sum_{\mathbf{i}}\Big(\hat{n}_{\mathbf{i}\uparrow}\hat{n}_{\mathbf{i}\downarrow} - \frac{\hat{n}_{\mathbf{i}\uparrow}+\hat{n}_{\mathbf{i}\downarrow}}{2}\Big)
 +\mu\sum_{\mathbf{i}\sigma}\hat{n}_{\mathbf{i}\sigma},
\end{aligned}\end{equation}
with $\varepsilon_{\mathbf{k}}=-2t(\cos k_x+\cos k_y)$, $\mathcal{L}_{\mathbf{k}}=2\lambda(\sin k_y-i\sin k_x)$, and $\hat{n}_{\mathbf{i}\sigma}=c_{\mathbf{i}\sigma}^{\dag}c_{\mathbf{i}\sigma}$ representing the density operator with spin $\sigma=\uparrow,\downarrow$ on the lattice site $\mathbf{i}=(i_x,i_y)$. The momentum $k_x$ and $k_y$ are defined in units of $2\pi/L$ with the system size $N_s=L^2$. We denote the fermion filling as $n=N/N_s$ with $N$ as the total number of fermions in the system. The nearest-neighbor hopping $t$, on-site Coulomb interaction $U$ ($<0$), the SOC strength $\lambda$, and chemical potential $\mu$ are model parameters. Within the above formulation, the system is at half filling with $n=1$ for $\mu=0$ due to the particle-hole symmetry~\cite{Tang2014}, and it is hole doped for $\mu>0$. Throughout this work, we set $t$ as the energy scale, and we focus mostly on the doped systems with the fermion filling $n<1$. 

The previous study~\cite{Rosenberg2017} showed that the model in Eq.~(\ref{eq:2DHamlt}) has a supersolid ground state with coexisting charge and superfluid long-range orders at half filling. Away from this special point, the superfluidity survives for arbitrary filling with arbitrary interaction strength~\cite{Shi2016}. Since the SOC term breaks the spin SU(2) symmetry and results in two helical bands for noninteracting case, the corresponding superfluid state with interaction is composed of both spin-singlet and triplet Cooper pairs, whose pairing properties are the main content of this work.

We then apply the finite-temperature AFQMC algorithm~\cite{Blankenbecler1981,Hirsch1983,White1989,Yuanyao2019} to numerically solve the lattice model in Eq.~(\ref{eq:2DHamlt}). It is free of fermion sign problem at arbitrary filling due to the time-reversal symmetry~\cite{Congjun2005}. The scheme of the AFQMC method is first to decouple the two-body interactions into free fermions coupled with auxiliary fields and then to calculate the fermionic observables through importance sampling of the field configurations. Practically, the imaginary-time discretization of the inverse temperature as $\beta=M\Delta\tau$, the symmetric Trotter-Suzuki decomposition $e^{-\Delta\tau\hat{H}}=e^{-\Delta\tau\hat{H}_0/2}e^{-\Delta\tau\hat{H}_I}e^{-\Delta\tau\hat{H}_0/2}+\mathcal{O}[(\Delta\tau)^3]$ (with $\hat{H}_0$ and $\hat{H}_I$ as the free and interaction parts of the Hamiltonian), and the Hubbard-Stratonovich (HS) transformation are successively implemented. The discrete HS transformation with the spin-$\hat{s}_z$ decomposition rather than the usual charge channel~\cite{Hirsch1983} is adopted for the attractive $U$ interaction to suppress the fluctuations of pairing related observables. Other algorithmic advances and techniques applied here include the Fast Fourier Transform (FFT) between the real and momentum space~\cite{Yuanyao2019}, the delayed version of local update~\cite{McDaniel2017}, and the $\tau$-line type of global update~\cite{Scalettar1991}, which together improve the efficiency of the numerical simulations. For further details of the AFQMC algorithm, we refer to the reviews in Ref.~\cite{Assaad2008,Chang2015}.

\section{Numerical results}
\label{sec:QMCresults}

In this section, we present the AFQMC simulation results of the lattice model in Eq.~(\ref{eq:2DHamlt}), including the pairing structure, the pairing correlation functions and BKT transition. Our AFQMC calculations reach the linear system size $L=20$ with the temperature as low as $T/t=0.025$ to sufficiently access the superfluidity (quasi-long-range ordered or quasi-condensate) regime. We mainly concentrate on the pairing properties away from half filling, for which the charge density wave does not have long-range order (see details in Appendix~\ref{sec:AppendixA}). The parameter $\Delta\tau t=0.05$ is chosen mostly in this work, which has been tested to safely eliminate the Trotter error, except for the strong interactions where smaller $\Delta\tau$ is applied. Periodic boundary conditions in both directions are applied for all the calculations. 

\subsection{Condensate fractions and pair wave functions}
\label{sec:CondnstPairWvfc}

The contributions of spin-singlet and triplet channels to the fermion pairing can be quantified by the corresponding condensate fractions~\cite{Anna2011,*Anna2012}. On the other hand, properties of the Cooper pairs, including their sizes and the fermion momentum, can be obtained from the pair wave functions~\cite{Shi2016,Rosenberg2017}. The computation of these quantities involves the following pairing matrix in momentum space~\cite{Shi2016,Rosenberg2017} as 
\begin{equation}\begin{aligned}
\label{eq:PairMat}
M(\mathbf{k},\ell;\mathbf{k}^{\prime},\ell^{\prime}) = \langle \Delta_{\ell}^{\dag}(\mathbf{k}) \Delta_{\ell^{\prime}}(\mathbf{k}^{\prime}) \rangle,
\end{aligned}\end{equation}
with $\ell = s$ or $t_{\uparrow}$ or $t_{\downarrow}$,  and $\Delta_{\ell}^{\dag}(\mathbf{k}) $ as spin-singlet and triplet pairing operators with {\it zero center-of-mass momentum} as
\begin{equation}\begin{aligned}
&\hspace{0.6cm}
\Delta_{s}^{\dag}(\mathbf{k}) = 
\frac{1}{\sqrt{2}}(c_{\mathbf{k}\uparrow}^{\dag}c_{-\mathbf{k}\downarrow}^{\dag}-c_{\mathbf{k}\downarrow}^{\dag}c_{-\mathbf{k}\uparrow}^{\dag})\\
&\Delta_{t_{\uparrow}}^{\dag}(\mathbf{k}) = c_{\mathbf{k}\uparrow}^{\dag}c_{-\mathbf{k}\uparrow}^{\dag}
\hspace{0.8cm}
\Delta_{t_{\downarrow}}^{\dag}(\mathbf{k}) = c_{\mathbf{k}\downarrow}^{\dag}c_{-\mathbf{k}\downarrow}^{\dag}.
\end{aligned}\end{equation}
Note that the third component of the triplet pairing eliminates by symmetry. In $N_s=L^2$ finite system, the pairing matrix $\bf{M}$ is a $3N_s\times 3N_s$ matrix. Attributed to the FFT algorithm applied in our numerical calculations, we can compute the equal-time, momentum-space single-particle Green's function matrix $\mathbf{G}=\{G_{\mathbf{k}\sigma,\mathbf{k}^{\prime}\sigma^\prime}=\langle c_{\mathbf{k}\sigma}c_{\mathbf{k}^{\prime}\sigma^\prime}^{\dagger}\rangle_{\tau}\}$, and thus directly measure the pairing matrix defined in Eq.~(\ref{eq:PairMat}) for a single auxiliary-field configuration via Wick decomposition.

\begin{figure}[t]
\centering
\includegraphics[width=0.99\columnwidth]{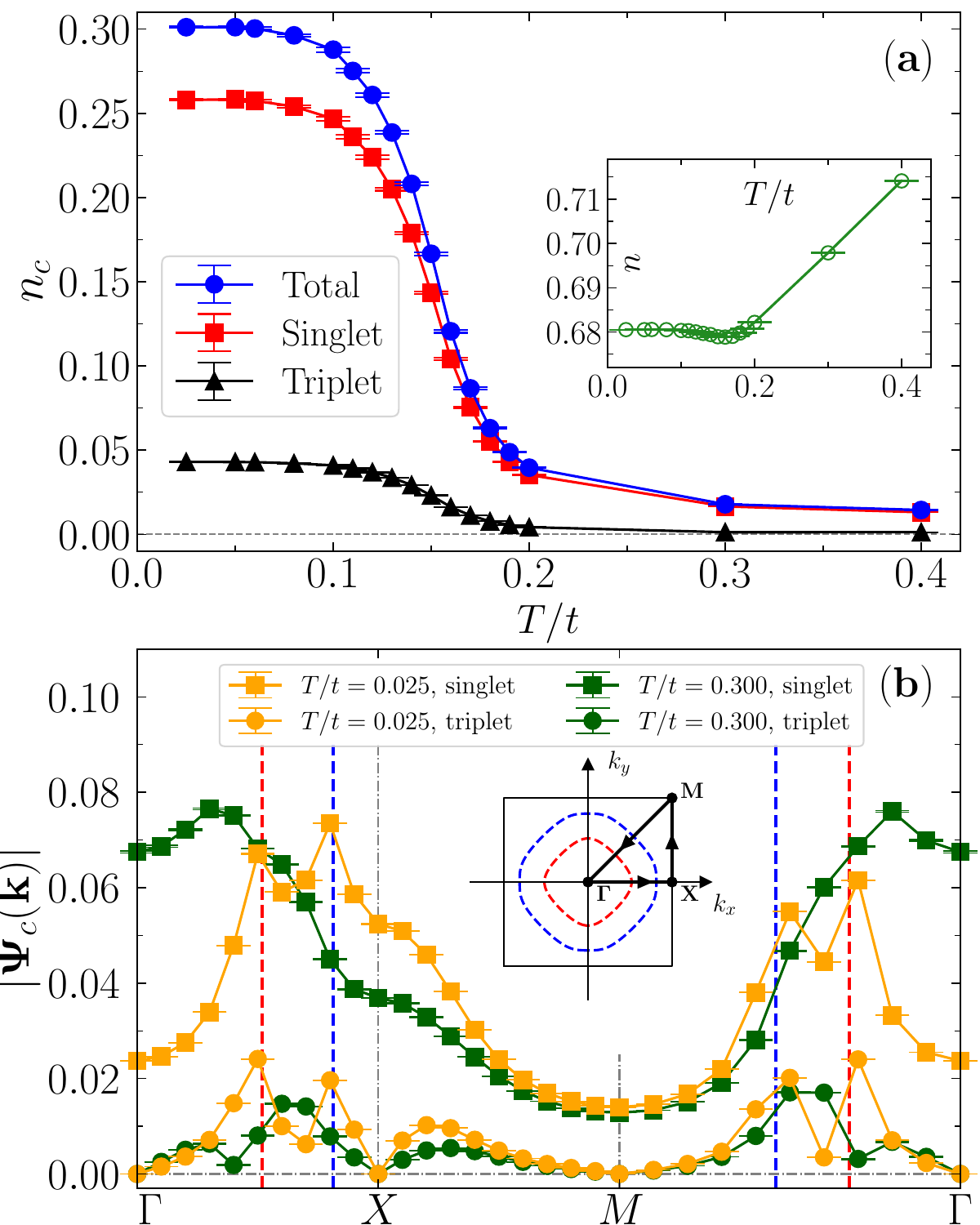}
\caption{\label{fig:Fig01CondWvfc} Illustration of the condensate fractions and pair wave functions in momentum space. Plotted in panel (a) are the total, singlet and triplet condensate fractions versus the temperature. The inset is the corresponding fermion filling. Panel (b) presents the magnitudes of momentum-space pair wave functions $|\Psi_c(\mathbf{k})|$ in both singlet and triplet channels for two temperatures $T/t=0.025$ and $0.30$. The inset plots the Fermi surfaces (red and blue dashed lines) for the fermion filling of $T/t=0.025$ case and the high-symmetry path (black solid lines with arrows) with $\boldsymbol{\Gamma}$, $\mathbf{X}$ and $\mathbf{M}$ points in Brillouin zone. These calculations are performed for $L=20$ system with $U/t=-4$, $\lambda/t=0.5$ and $\mu/t=0.5$.}
\end{figure}

Then from the leading eigenvalue $N_c$ of the pairing matrix, we can obtain the total condensate fraction as $n_c=N_c/(N/2)$~\cite{Yang1962,Boronat2005}. The corresponding eigenstate of $N_c$ is the momentum-space pair wave function $\boldsymbol{\Psi}_c$, which consists of the singlet and triplet components as $\boldsymbol{\Psi}_c = (\boldsymbol{\Psi}_{c,s},\boldsymbol{\Psi}_{c,t_{\uparrow}},\boldsymbol{\Psi}_{c,t_{\downarrow}})^{\rm T}$ with every component as a $N_s$-dimensional vector. For the lattice model in Eq.~(\ref{eq:2DHamlt}), the two triplet channels are degenerate as $\boldsymbol{\Psi}_{c,t_{\uparrow}}=\boldsymbol{\Psi}_{c,t_{\downarrow}}$ due to the spin-inversion symmetry. Thus, we define the overall triplet pair wave function as $\boldsymbol{\Psi}_{c,t}=\sqrt{2}\boldsymbol{\Psi}_{c,t_{\uparrow}}$. Then within normalized $\boldsymbol{\Psi}_c$, we assign the condensate fractions of spin-singlet and triplet pairing as $n_{c,s}=n_c\times(\boldsymbol{\Psi}_{c,s}\boldsymbol{\Psi}_{c,s}^{\rm T})$ and $n_{c,s}=n_c\times(\boldsymbol{\Psi}_{c,t}\boldsymbol{\Psi}_{c,t}^{\rm T})$, respectively. Thus, the relation $n_c=n_{c,s}+n_{c,t}$ obviously holds, with $n_{c,s}/n_c$ and $n_{c,t}/n_c$ as the contributions of singlet and triplet channels to the pairing. The square $|\boldsymbol{\Psi}_{c,\ell}({\bf{k}})|^2$ ($\ell = s,t$) stands for the probability of fermions with momentum $\bf{k}$ participating the pairing. We can further obtain the corresponding real-space pair wave functions $\psi_{c,s}(\bf{r})$ and $\psi_{c,t}(\bf{r})$ by Fourier transform of $\boldsymbol{\Psi}_{c,s}$ and $\boldsymbol{\Psi}_{c,t}$. Similarly, $|\psi_{s}({\bf{r}})|^2$ and $|\psi_{t}({\bf{r}})|^2$ represent probabilities of spin-singlet and triplet Cooper pairs with distance $\bf{r}$ of the two fermions, and they actually reflect the size of the pairs. 

We note that the finite-temperature condensate fraction should vanish in the thermodynamic limit for both the quasi-condensate and disordered phases. Alternatively, its finite-size results can present plentiful information about the pairing properties~\cite{Shi2016,Rosenberg2017,Yuanyao2022}. Attributed to the distinct behaviors of pairing correlation accross the BKT transition, the finite-size condensate fraction should also undergo a qualitative change around the transition, which serves as an effective tool to locate the BKT transition temperature~\cite{Yuanyao2022}. Thus, throughout this work, we always deal with finite-size AFQMC results of condensate fraction at finite temperatures.

As a demonstration, we show the typical results of condensate fractions and pair wave functions for a specific group of parameters in Fig.~\ref{fig:Fig01CondWvfc}. With lowering temperature, both the spin-singlet and triplet condensate fractions monotonically increases from the high-temperature normal state to the low-temperature superfluid phase, and then saturates to the ground-state values as indicated by the plateau achieved with $T/t\le 0.06$ results. They also exhibit a rapid increase at a specific temperature, for which the BKT transition should be responsible (see Sec.~\ref{sec:ComputeBKT}). Remarkably, the triplet channel has a rather small contribution to the total condensate fraction (less than $15\%$ approaching $T=0$). As shown in Fig.~\ref{fig:Fig01CondWvfc}(b), the momentum-space pair wave functions of both singlet and triplet channels have peaks around the Fermi surfaces of the two helical bands for the intermediate interaction $U/t=-4$ at low temperature, which resembles the results without SOC~\cite{Yuanyao2019,Yuanyao2022}. This is also consistent with the fundamental picture of BCS theory that the fermions around the Fermi surfaces dominates the pairing in the weakly interacting regime~\cite{Cooper1956}. In contrast, the high-temperature results of the pair wave functions seem featureless as the system is in the normal state. Note that the node at $\boldsymbol{\Gamma}$ point in triplet pair wave function indicates its antisymmetry, while the singlet component is symmetric without node.

\subsection{The mixed-parity pairing structure}
\label{sec:PairStruct}

Based on the results and discussions in Sec.~\ref{sec:CondnstPairWvfc}, we then concentrate on the mixed-parity pairing structure for physically relevant parameter regimes, revealed by the numerical results of condensate fractions and pair wave functions. tunning parameters including the temperature, interaction strength, SOC and the chemical potential are accounted for in our AFQMC simulations. 

The condensate fractions versus the temperature typically shares similar behavior as the results shown in Fig.~\ref{fig:Fig01CondWvfc}(a), with differences lying in the specific numbers and BKT transition temperatures. With varying interaction strength, SOC and chemical potential, we perform the AFQMC simulations with $T/t=0.10$, at which most systems studied as follows fall into the superfluid phase and the condensate fractions are close to the corresponding $T=0$ results [similar to Fig.~\ref{fig:Fig01CondWvfc}(a)]. 

\begin{figure}[t]
\centering
\includegraphics[width=0.97\columnwidth]{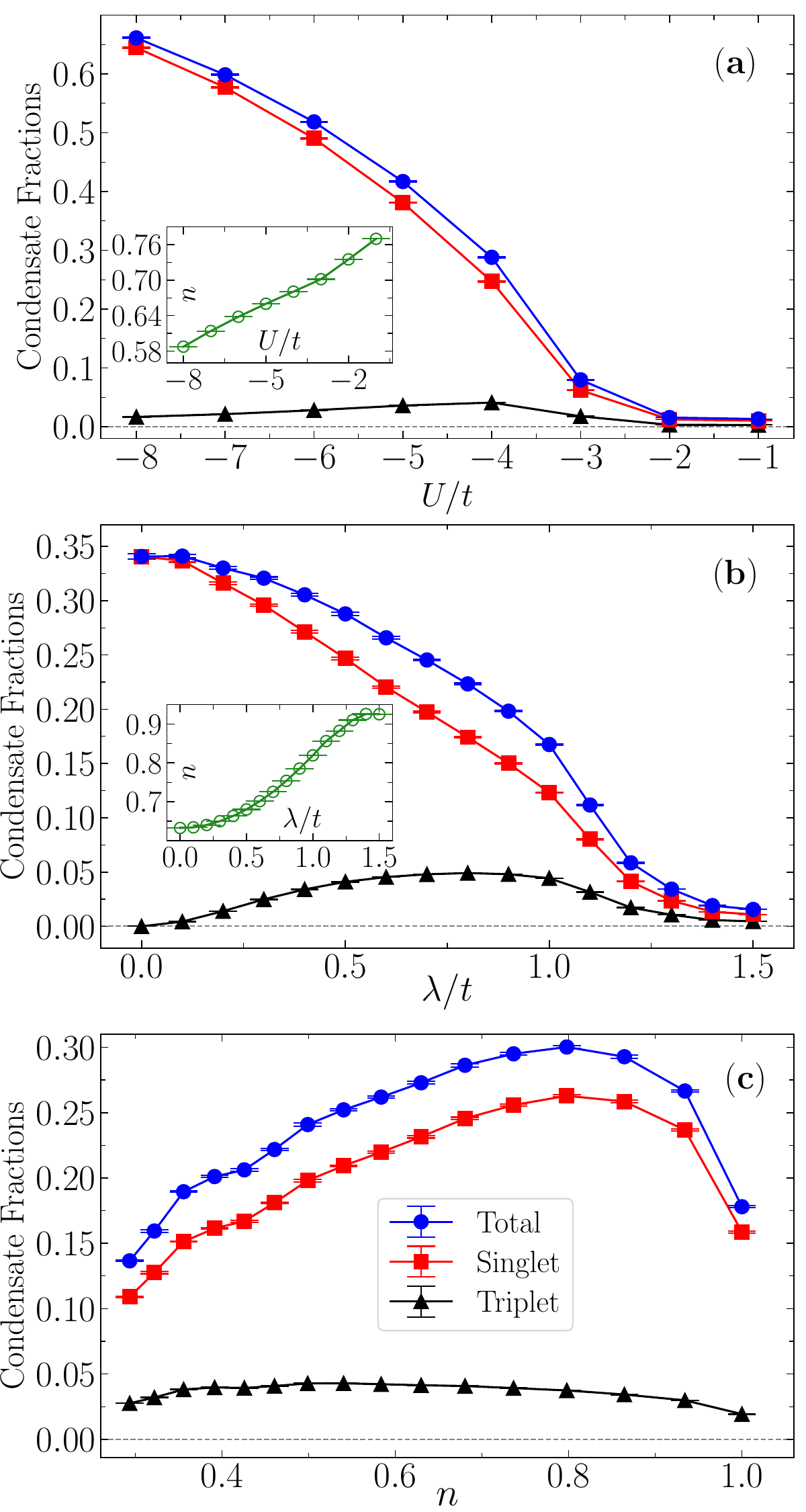}
\caption{\label{fig:Fig02CondFrac} The condensate fractions $n_c$, $n_{c,s}$ and $n_{c,t}$ versus various tunning parameters. (a) Tune interaction strength $U/t$ with $\lambda/t=0.5$ and $\mu/t=0.5$; (b) Tune SOC $\lambda/t$ with $U/t=-4$ and $\mu/t=0.5$; (c) Tune the fermion filling $n$ by chemical potential $\mu/t$ with $U/t=-4$ and $\lambda/t=0.5$. The insets in Panel (a) and (b) plot the results of corresponding filling. These calculations are performed for $L=20$ system with temperature $T/t=0.1$.}
\end{figure}

\begin{figure*}[t]
	\centering
	\includegraphics[width=1.75\columnwidth]{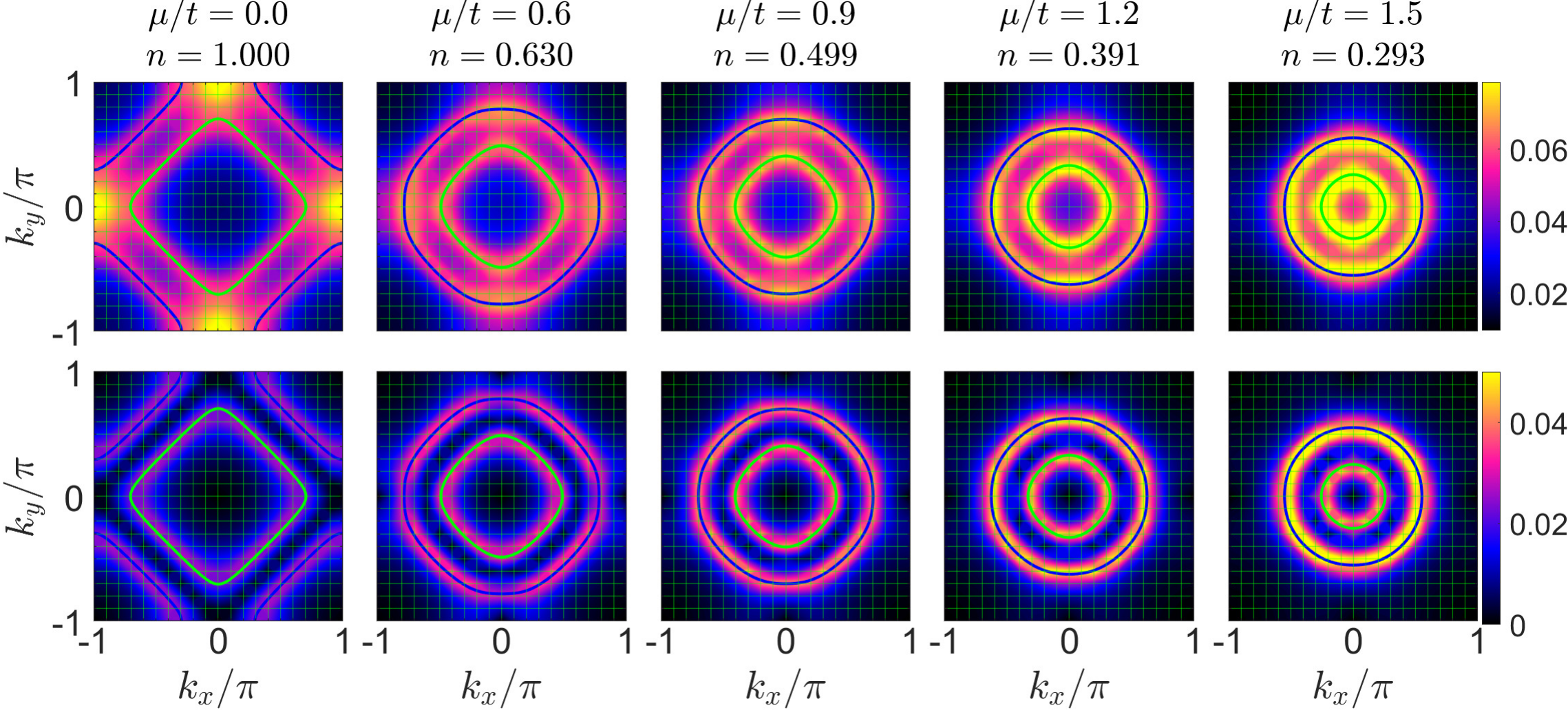}
	\caption{\label{fig:Fig03PairWfVsMu} The singlet (top) and triplet (bottom) pair wave functions in momentum space, $|\boldsymbol{\Psi}_{c,s}({\bf{k}})|$ and $|\boldsymbol{\Psi}_{c,t}({\bf{k}})|$, versus chemical potential $\mu/t$ with corresponding fermion filling $n$  (error bars are in the fourth/fifth digits and are thus neglected) shown on top of the plots. The error bars of $n$ are in the fourth or fifth digits and thus are neglected. The noninteracting Fermi surfaces at $T=0$ of the two helical bands are also plotted with the green and blue solid lines. These calculations are performed for $L=20$ system with $T/t=0.10$ and $U/t=-4$, $\lambda/t=0.5$.}
\end{figure*}

\begin{figure*}
	\centering
	\includegraphics[width=1.75\columnwidth]{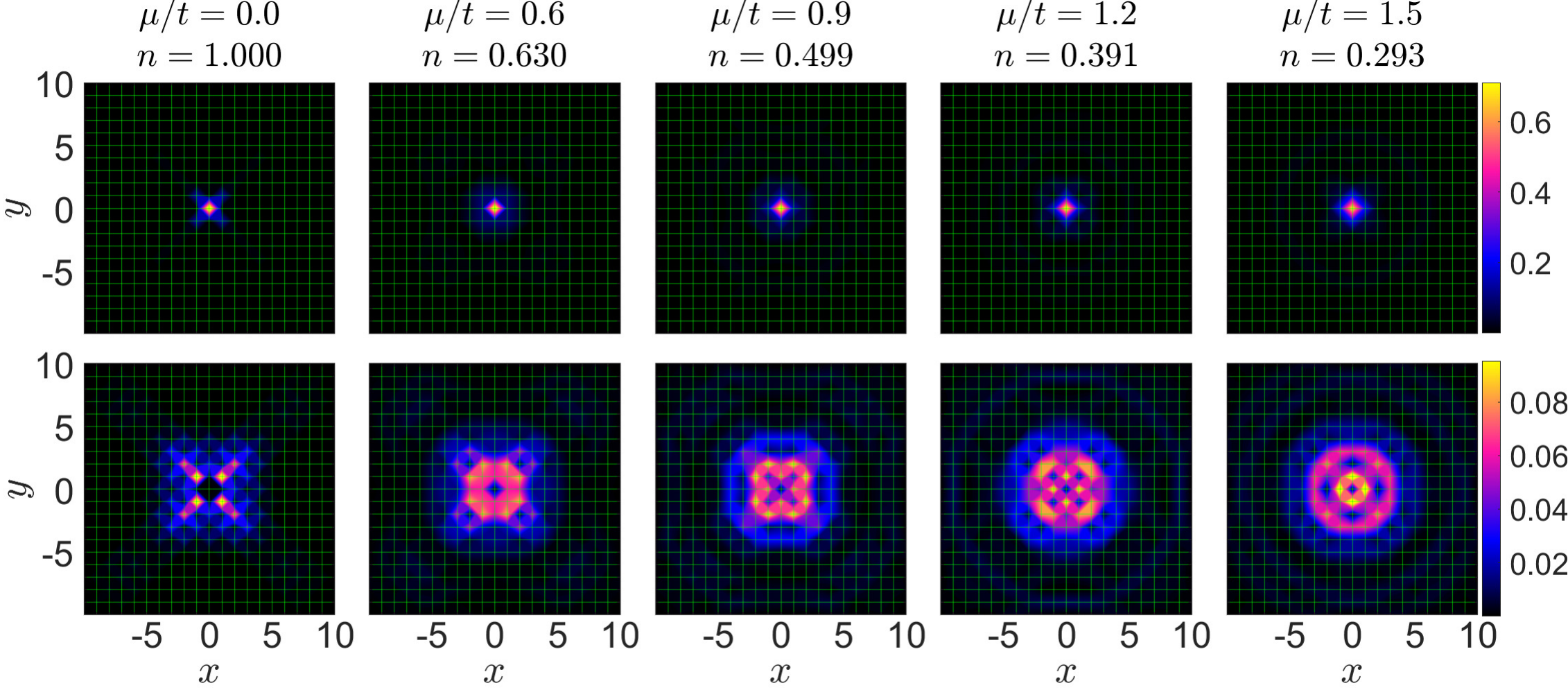}
	\caption{\label{fig:Fig04PairWfVsMu} The singlet (top) and triplet (bottom) pair wave functions in real space, $|\psi_{s}({\bf{r}})|$ and $|\psi_{t}({\bf{r}})|$, versus chemical potential $\mu/t$. Simulation parameters are the same as Fig.~\ref{fig:Fig03PairWfVsMu}.}
\end{figure*}

In Fig.~\ref{fig:Fig02CondFrac}, we present one of the key results in this work, as condensate fractions with tunning parameters other than the temperature. First, with increasing on-site interaction, the system hosts the BCS-BEC crossover from extended Cooper pairs to tightly bounded molecules~\cite{Giancarlo2018}. Fig.~\ref{fig:Fig02CondFrac}(a) shows that the singlet condensate fraction simply increases during the crossover, while the triplet contribution has a peak around $U/t=-4$ (for $\lambda/t=0.5$). This difference can be understood as follows. Turning on the interaction can first enhance the pairing in both channels as well as the condensate fractions. Cross some intermediate $U$, the interaction begins to frustrate the triplet pair formation and continues to increase the singlet pairs, due to the nature of the attraction between fermions with unlike spins. The results suggest that the triplet contribution to the pairing is most significant [$\sim 21\%$ as in Fig.~\ref{fig:Fig02CondFrac}(a)] in the intermediate interaction regime, whose specific value of $U/t$ should depend on SOC strength. These results are qualitatively consistent with those from the ground-state calculations of 2D spin-orbit-coupled Fermi gas~\cite{Shi2016}. Then, with tunning SOC strength, the condensate fractions for $U/t=-4$ are plotted in Fig.~\ref{fig:Fig02CondFrac}(b). The decreasing of singlet condensate fraction with $\lambda/t$ can be explained by the enlarged bandwidth $W(t,\lambda)$~\cite{Tang2014} and the reduced effective interaction $U/W$. However, the triplet condensate fraction is first enhanced by SOC, due to the fact that SOC is the essential source of triplet pairing in presence of Hubbard interaction. Then the effect from reduced $U/W$ sets in, and the competition results in a broaden peak around $\lambda/t=0.5\sim1.0$. The biggest contribution from the triplet channel to the pairing is $\sim 30\%$ around $\lambda/t=1.3$, where nevertheless the total condensate fraction is only $0.043$. Finally, in Fig.~\ref{fig:Fig02CondFrac}(c), we show the numerical results versus the fermion filling (by tunning the chemical potential). Both the singlet and total condensate fractions reach the maximum around $n=0.80$, while the triplet one possesses a wide plateau regarding the filling. The triplet contribution saturates to largest value $\sim 20\%$ towards the low filling regime. As discussed above, for the simulation temperature $T/t=0.10$, the system evolves from the normal state at half filling (with $\mu=0$) to the superfluid phase with increasing doping. Thus, the results in Fig.~\ref{fig:Fig02CondFrac}(c) might indicate that the maximal BKT transition temperature is achieved around the filling $n=0.80$~\cite{Tang2014}. Combining all the results in Fig.~\ref{fig:Fig02CondFrac}, we can conclude that the spin-singlet pairing always has the predominant contribution than the triplet channel to the mixed-parity pairing in the system.

\begin{figure}[t]
\centering
\includegraphics[width=0.99\columnwidth]{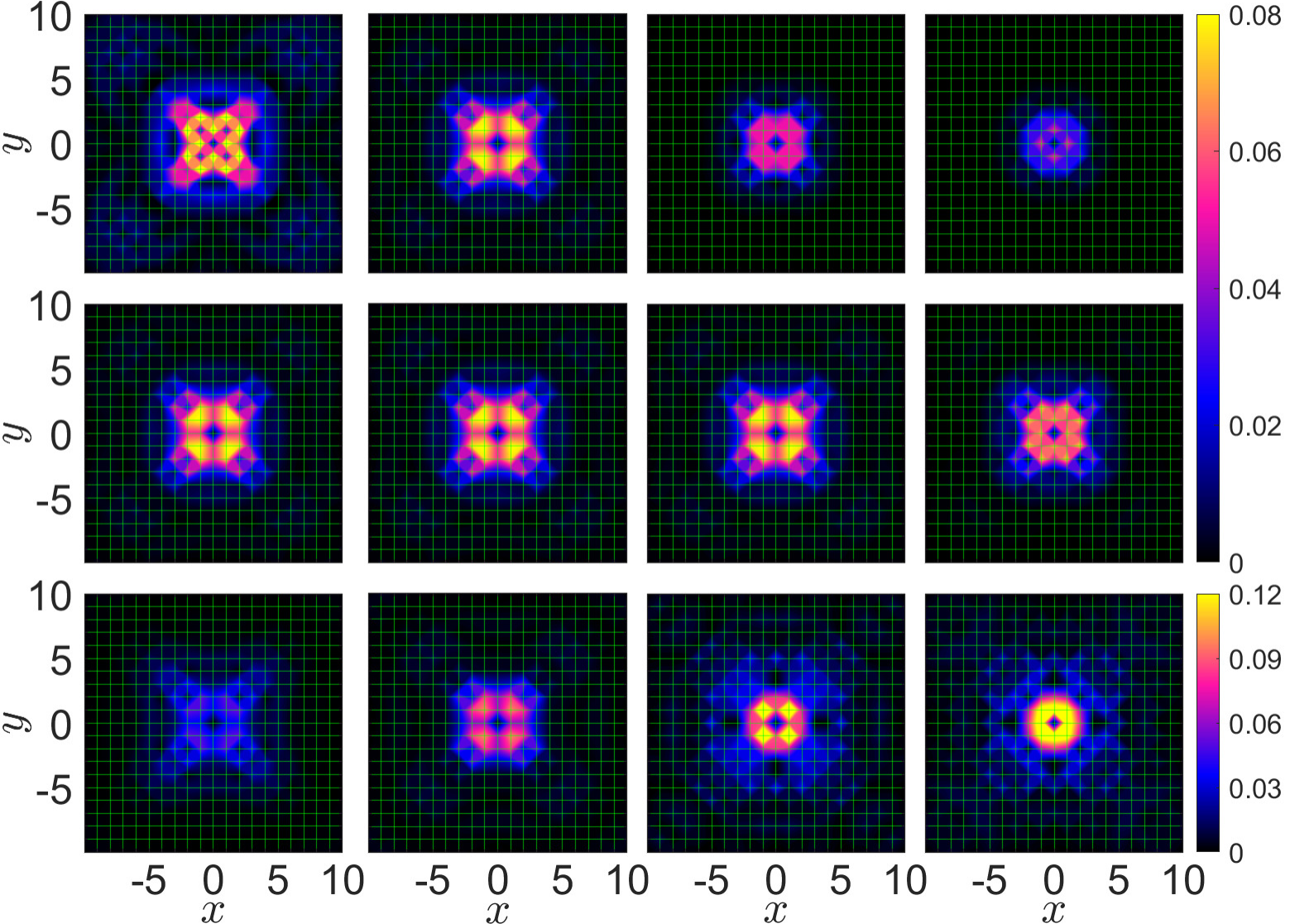}
\caption{\label{fig:Fig05PairWfVsParamt} The triplet pair wave function in real space $|\psi_{t}({\bf{r}})|$ versus (top row) interaction strength $-U/t=2,4,6,8$ with $T/t=0.10,\lambda/t=0.5,\mu/t=0.5$; (middle row) the temperature $T/t=0.025,0.05,0.14,0.20$ with $U/t=-4,\lambda/t=0.5,\mu/t=0.5$; (bottom row) SOC strength $\lambda/t=0.3,0.6,1.2,1.5$ with $T/t=0.10,U/t=-4,\mu/t=0.5$. These calculations are performed for $L=20$ system.}
\end{figure}

\begin{figure}[t]
\centering
\includegraphics[width=0.95\columnwidth]{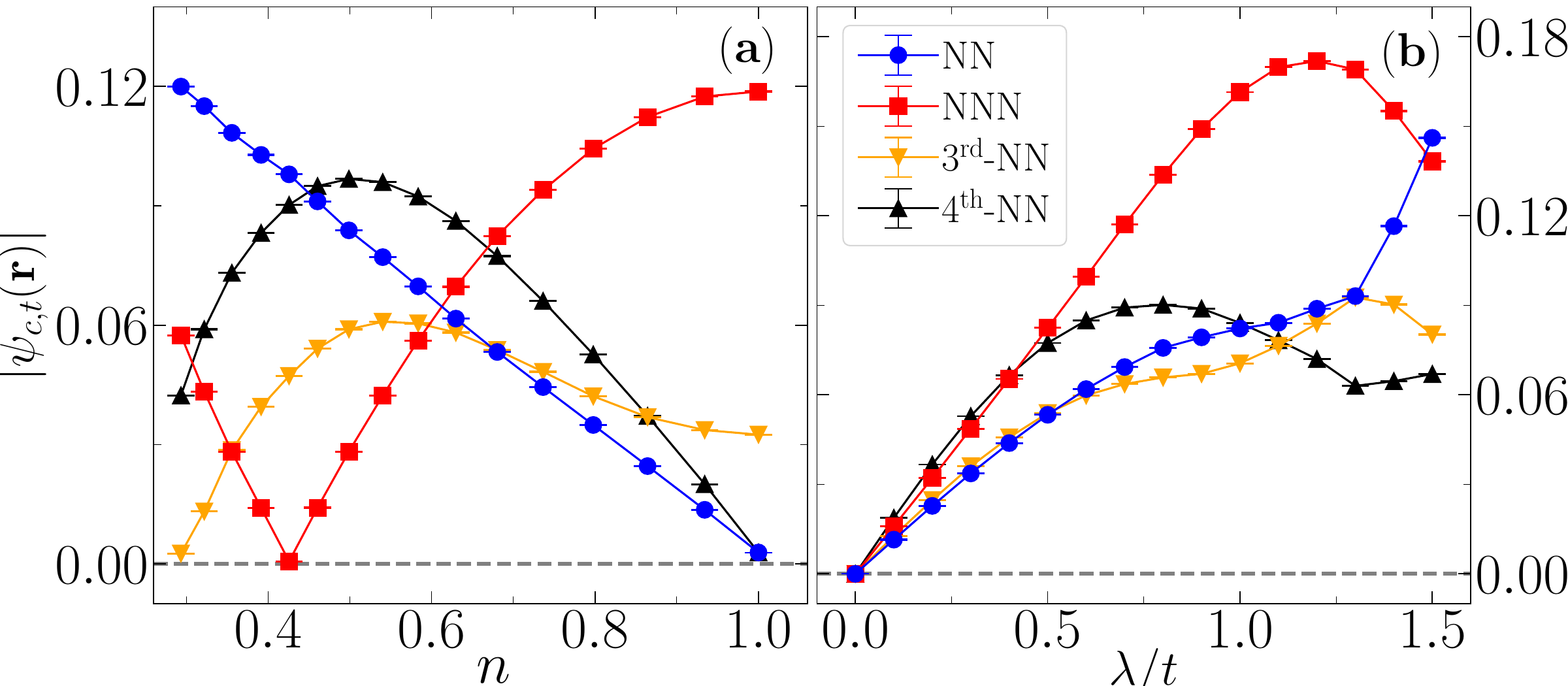}
\caption{\label{fig:Fig06TripletNN} The amplitudes of real-space triplet pair wave function $|\psi_{c,t}(\mathbf{r})|$ with the distance $r$ equal to NN, NNN, $3^{\rm rd}$-NN and $4^{\rm th}$-NN sites, with (a) tunning the fermion filling $n$ and (b) varying the SOC strength. The other simulation parameters for Panel (a) and (b) are the same as Fig.~\ref{fig:Fig04PairWfVsMu} and the bottom row of Fig.~\ref{fig:Fig05PairWfVsParamt}, respectively. These calculations are performed for $L=20$ system.}
\end{figure}

We then turn to the results of the pair wave functions. First, their evolutions versus the chemical potential in momentum space and real space are illustrated in Fig.~\ref{fig:Fig03PairWfVsMu} and Fig.~\ref{fig:Fig04PairWfVsMu}, respectively. For half filling, our results are quantitatively consistent with the $T=0$ results in Ref.~\cite{Rosenberg2017}. Increasing the chemical potential results in smaller fermion filling, and the corresponding noninteracting Fermi surfaces at $T=0$ of the two helical bands (dashed lines in Fig.~\ref{fig:Fig03PairWfVsMu}), which are determined from the corresponding fermion filling from finite-T AFQMC calculation, shrinks towards circles. It's clear that for the intermediate interaction $U/t=-4$ the pair wave functions in both channels show sharp peaks in the vicinity of the Fermi surfaces, regardless of the filling. With the increasing interaction, the results should gradually become smooth in the whole Brillouin zone (not shown) without apparent peaks~\cite{Shi2016} indicating the deviation from BCS theory. In contrast, the singlet and triplet pair wave functions in real space show significant difference, as shown in Fig.~\ref{fig:Fig04PairWfVsMu}. The localized peaks in singlet pair wave function $|\psi_{s}({\bf{r}})|$ clearly shows that the singlet pairing mainly has a local origin with on-site pairs. However, the Pauli principle prohibits such on-site triplet pair formation, resulting in zero value at $\bf{r}=0$. Instead, the triplet pair wave function $|\psi_{t}({\bf{r}})|$ is more extended and has very rich patterns and evolutions along with decreasing fermion filling. Multi-peak structures appear in $|\psi_{t}({\bf{r}})|$, with the largest amplitude locations changing from the next-nearest-neighbor (NNN) sites at half filling, to intermediate fourth-nearest-neighbor ($4^{\rm th}$-NN), and finally to the nearest-neighbor (NN) sites at low filling, as shown in Fig.~\ref{fig:Fig06TripletNN}(a). These finite-size triplet Cooper pairs within several NN sites can be explained by the real-space nature of Rashba SOC, which is actually NN spin-flip hopping. As a result, successive SOC hops can enhance the possibility to find another fermion at neighboring sites with the same spin as the one located at origin. Moreover, towards smaller filling, both of $|\psi_{s}({\bf{r}})|$ and $|\psi_{t}({\bf{r}})|$ show more extended behaviors due to the enlarged wavelength $\sim 2\pi/k_F$. 

As for the other tunning parameters, the pair wave functions in momentum space show similar behaviors as illustrated in Fig.~\ref{fig:Fig03PairWfVsMu}, and in real space for singlet channel as $|\psi_{s}({\bf{r}})|$ are also dominated by the center peak as Fig.~\ref{fig:Fig04PairWfVsMu}. Thus, we now concentrate on triplet pair wave function in real space $|\psi_{t}({\bf{r}})|$ shown in Fig.~\ref{fig:Fig05PairWfVsParamt}. With increasing interaction strength [top row of Fig.~\ref{fig:Fig05PairWfVsParamt}], $|\psi_{t}({\bf{r}})|$ gradually evolves from rather extended pattern with multipeaks along diagonals, to local peaks at NN lattice sites, which illustrates the BCS-BEC crossover behavior in triplet pairing channel. With decreasing temperature [middle row of Fig.~\ref{fig:Fig05PairWfVsParamt}], the peaks in $|\psi_{t}({\bf{r}})|$ (located at NNN and $4^{\rm th}$-NN sites) simply become more significant and eventually stabilize, indicating entering the superfluid phase from the normal state. Increasing SOC strength, the $T=0$ AFQMC simulations at half filling~\cite{Rosenberg2017} showed a diamond pattern of $|\psi_{t}({\bf{r}})|$ with enhanced peak values at both NNN and $3^{\rm rd}$-NN sites. It behaves differently away from half filling [bottom row of Fig.~\ref{fig:Fig05PairWfVsParamt}]. As shown in Fig.~\ref{fig:Fig06TripletNN}(b), SOC first enhances all the finite-range triplet pairing for $\lambda/t<0.75$, where NNN and $4^{\rm th}$-NN pairing play the leading role. The NNN component is then further promoted by SOC, and NN and NNN pairing gradually becomes comparable towards large SOC, resulting in instead a square pattern as illustrated in Fig.~\ref{fig:Fig05PairWfVsParamt}. All the qualitative behaviors of pair wave functions results in Fig.~\ref{fig:Fig03PairWfVsMu}, Fig.~\ref{fig:Fig04PairWfVsMu} and Fig.~\ref{fig:Fig05PairWfVsParamt} do not change with the system size. 

%These complicated behaviors might be attribted to the competetion between enhanced NN spin-flip hops and the reduced band width as increasing SOC. Togather with the results in Fig.~\ref{fig:Fig04PairWfVsMu}, we can conclude that, for the physically relevant parameter regimes, the singlet channel is dominated by the local Cooper pairs while the triplet pairing is significantly more extended and mostly contributed by the NN and NNN fermion pairs. 

\subsection{The pairing correlation functions}
\label{sec:TripletCrFt}

In Sec.~\ref{sec:PairStruct}, the results in Fig.~\ref{fig:Fig02CondFrac} clearly present an optimal SOC strength and fermion filling regime where the triplet condensate fraction reaches the maximum. In this section, we pursue to understand this point from the aspect of the pairing correlation functions. 

\begin{figure}[b]
\centering
\includegraphics[width=0.92\columnwidth]{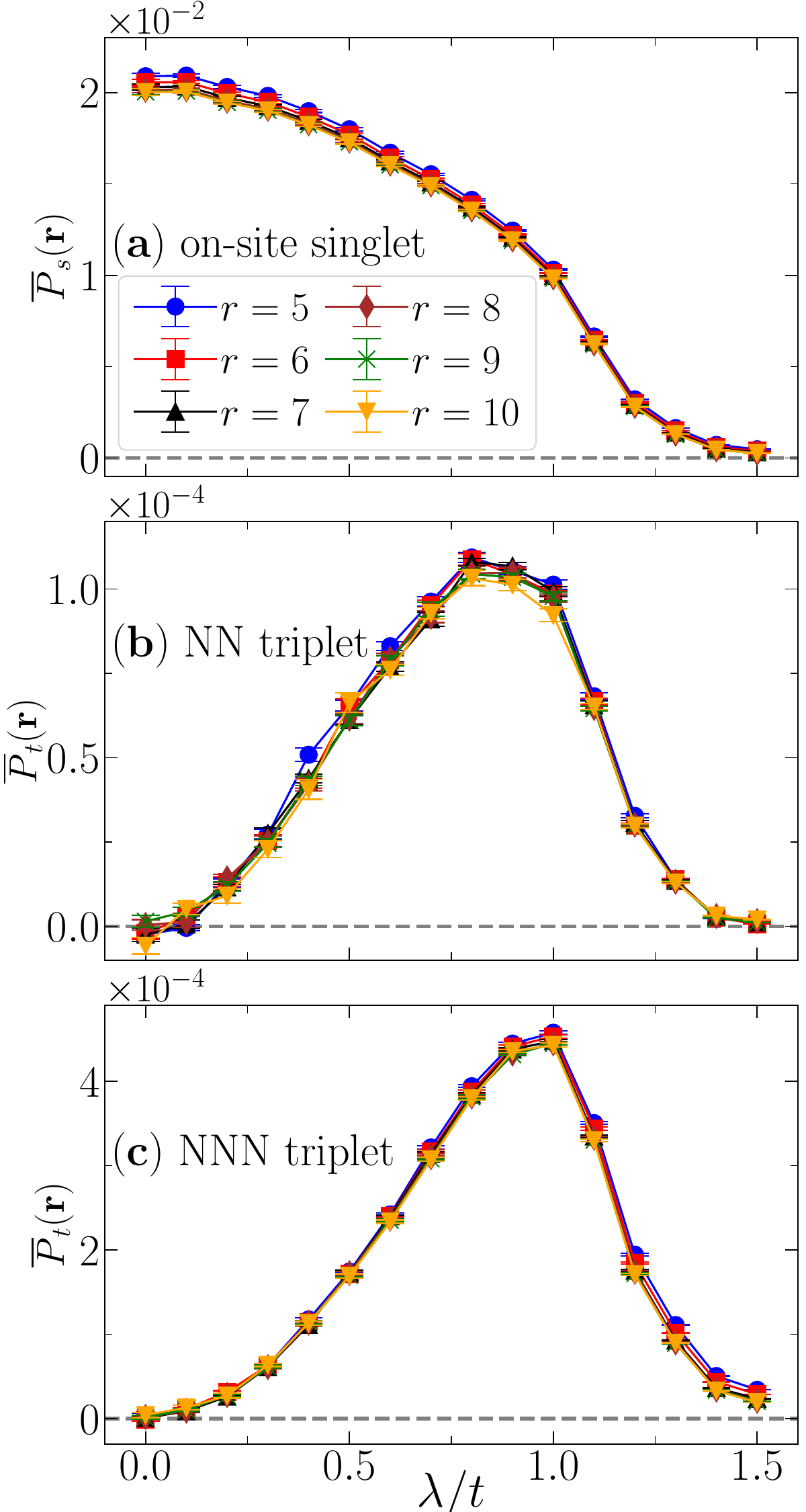}
\caption{\label{fig:Fig07CrFVsSOC} Vertex contribution of on-site singlet (top), NN triplet (middle) and NNN triplet (bottom) pairing correlation functions $\bar{P}_{\ell}(\mathbf{r})$ (with $\ell=s,t$) versus SOC strength. The correlations with distance $r=5\sim 10$ (along the $x$ axis) are plotted. These calculations are performed for $L=20$ system with $T/t=0.10$ and $U/t=-4$, $\mu/t=0.5$.}
\end{figure}

We define the real-space singlet and triplet pairing operators as
\begin{equation}\begin{aligned}
\label{eq:PairOp}
& \hat{\Delta}_{s,\bf{i}} = (c_{\bf{i}\uparrow}^{\dagger}c_{\bf{i}\downarrow}^{\dagger} + c_{\bf{i}\downarrow}c_{\bf{i}\uparrow}) / 2,\\
& \hat{\Delta}_{t,\bf{i}} = (c_{\bf{i}\uparrow}^{\dagger}c_{\bf{i}+\boldsymbol{\delta}\uparrow}^{\dagger} + c_{\bf{i}+\boldsymbol{\delta}\uparrow}c_{\bf{i}\uparrow}) / 2.
\end{aligned}\end{equation}
with $s$ and $t$ as singlet and triplet. For the triplet, we concentrate on NN and NNN pairing with $\boldsymbol{\delta}=(1,0)$ and $\boldsymbol{\delta}=(1,1)$ denoting the corresponding lattice vectors. We then measure the real space correlation functions $P_s(\mathbf{r})=\langle\hat{\Delta}_{s,\mathbf{i}}\hat{\Delta}_{s,\mathbf{i}+\mathbf{r}}\rangle$ and $P_t(\mathbf{r})=\langle\hat{\Delta}_{t,\mathbf{i}}\hat{\Delta}_{t,\mathbf{i}+\mathbf{r}}\rangle$, and the structure factors as their Fourier transformation $S_{\ell}(\mathbf{q})=\sum_{\mathbf{r}}P_{\ell}(\mathbf{r})e^{i\mathbf{q}\cdot\mathbf{r}}$ with $\ell=s$ or $t$. To directly evaluate the pure interaction contribution, we have also obtained the vertex contribution for the pairing correlations and structure factors, as $\bar{P}_s(\mathbf{r})$ and $\bar{S}_{\ell}(\mathbf{q})$, by subtracting the uncorrelated part~\cite{White1989b}. 

\begin{figure}[ht]
\centering
\includegraphics[width=0.92\columnwidth]{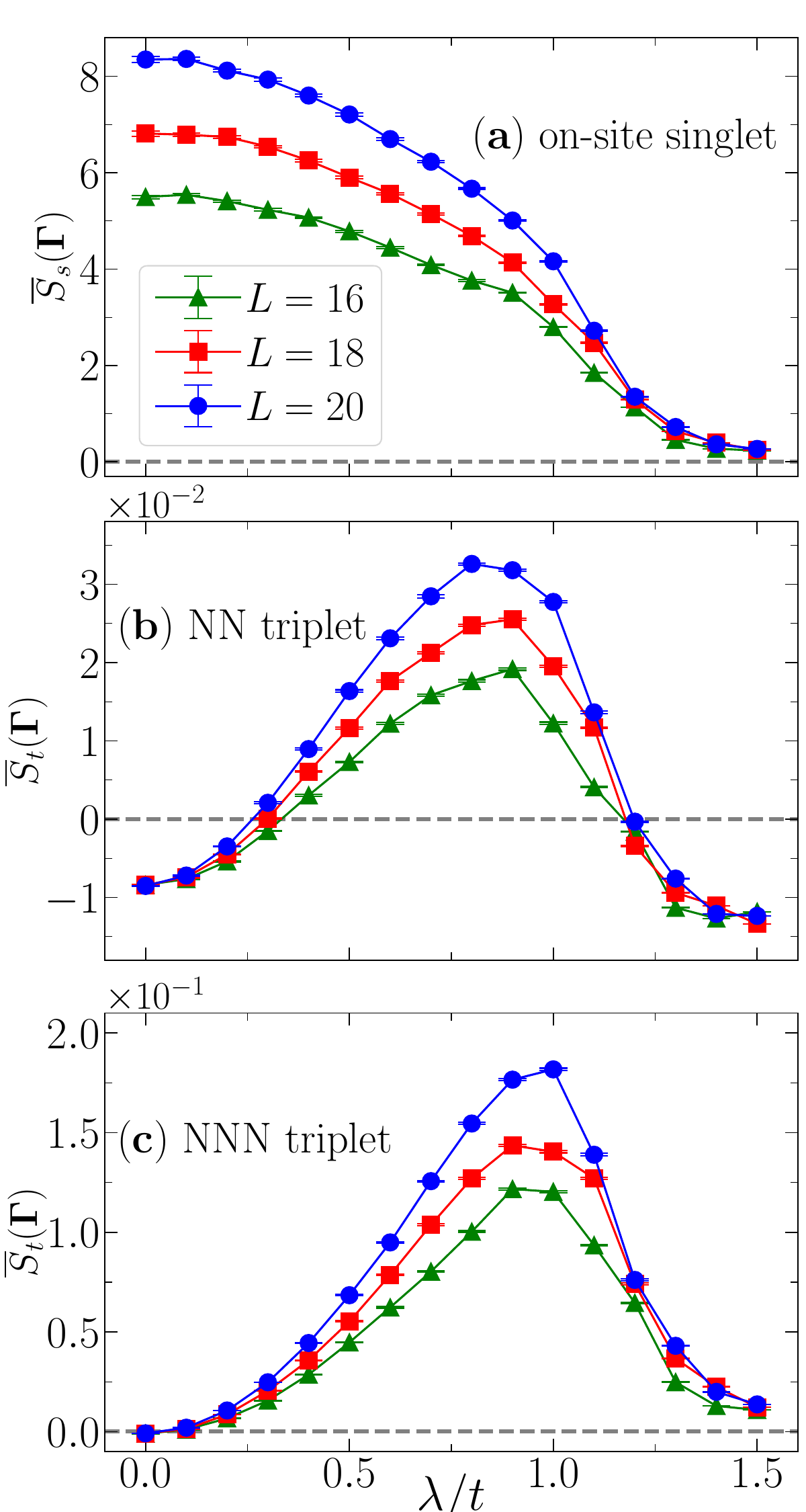}
\caption{\label{fig:Fig08StruVsSOC} Vertex contribution of on-site singlet (top), NN triplet (middle) and NNN triplet (bottom) pairing structure factors $\bar{S}_{\ell}(\mathbf{q}=\boldsymbol{\Gamma})$ (with $\ell=s,t$) versus SOC strength. These calculations are performed for $L=16,18,20$ systems with $T/t=0.10$ and $U/t=-4$, $\mu/t=0.5$.}
\end{figure}

In Fig.~\ref{fig:Fig07CrFVsSOC}, we present the vertex contributions to the pairing correlation functions of on-site singlet, NN and NNN triplet channels, with tunning SOC strength. All the positive vertex contributions to the correlations in Fig.~\ref{fig:Fig07CrFVsSOC} reveal that the on-site attractive interaction enhances the singlet and triplet pairing correlations with specific distances as $L/4\le r\le L/2$ (as $L=20$). These results contribute more than $90\%$ of the corresponding bare correlation functions (not shown). It's clear that SOC simply suppresses the singlet pairing correlation, while the NN and NNN triplet correlations show broaden peaks around $\lambda/t=0.8$ and $\lambda/t=1.0$. Moreover, the singlet correlation is stronger than the triplet ones around two orders of magnitude, indicating the dominant role of singlet channel. These results are in accordance with the behaviors of the corresponding condensate fractions shown in Fig.~\ref{fig:Fig02CondFrac}(b). The almost collapsed numerical data for different distances in Fig.~\ref{fig:Fig07CrFVsSOC} also highlight the superfluid phase of the system for the chosen parameters. Then, the vertex contributions of the pairing structure factors $\bar{S}_{\ell}(\mathbf{q}=\boldsymbol{\Gamma})$ (with $\ell=s,t$) with increasing SOC strength are illustrated in Fig.~\ref{fig:Fig08StruVsSOC}. They show similar behaviors as the real-space correlation functions. The negative vertex of $\bar{S}_{t}(\boldsymbol{\Gamma})$ with NN triplet for $\lambda/t<0.3$ and $\lambda/t>1.2$ means that the NN triplet pairing is not favored in these regimes. The growing numbers of $\bar{S}_{\ell}(\boldsymbol{\Gamma})$ in Fig.~\ref{fig:Fig08StruVsSOC} for all three quantities (especially in the intermediate SOC regime) with increasing system size also suggest quasi-long-range pairing orders. 

\begin{figure}[ht]
\centering
\includegraphics[width=0.99\columnwidth]{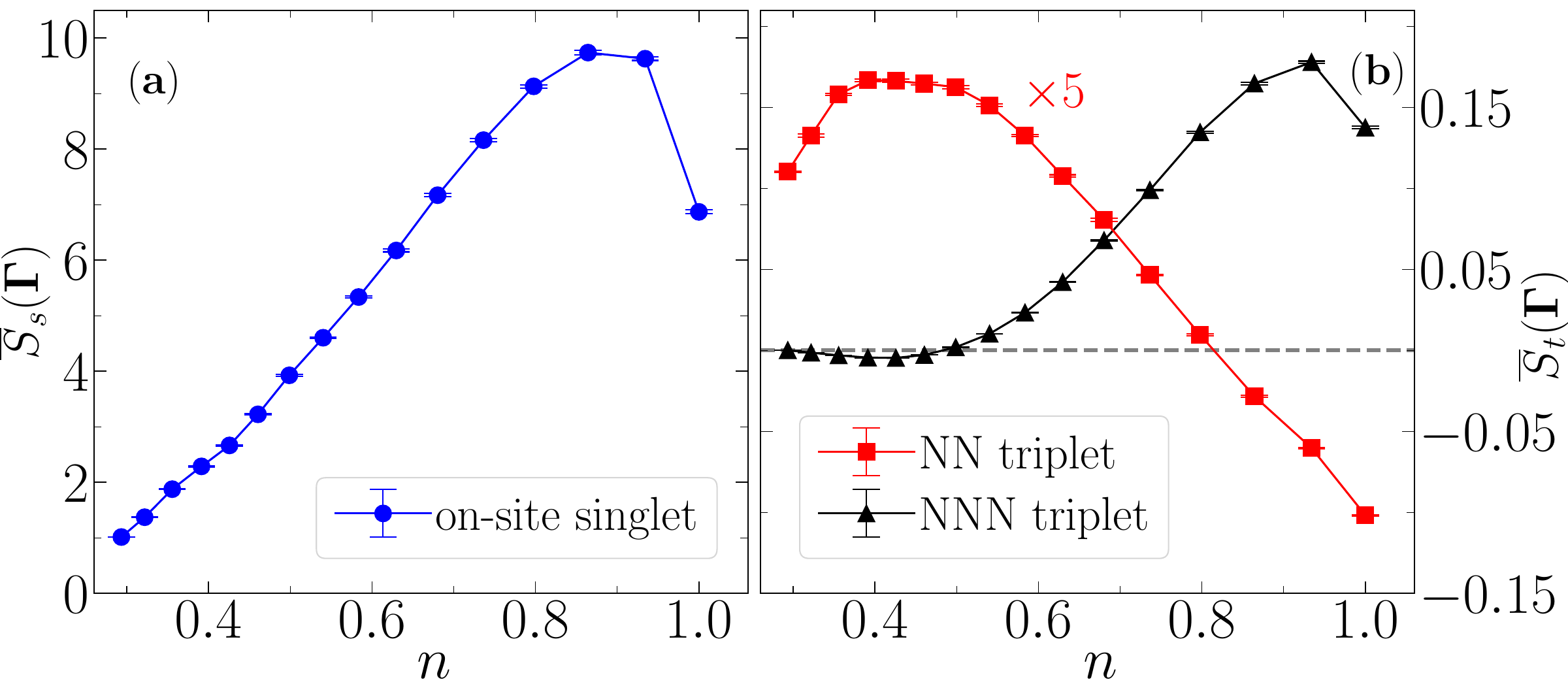}
\caption{\label{fig:Fig09StruVsMu} Vertex contribution of (a) on-site singlet, (b) NN triplet and NNN triplet pairing structure factors $\bar{S}_{\ell}(\boldsymbol{\Gamma})$ versus fermion filling. These calculations are performed for $L=20$ systems with $T/t=0.10$ and $U/t=-4$, $\lambda/t=0.5$.}
\end{figure}

Similarly, the results of condensate fractions versus the fermion filling in Fig.~\ref{fig:Fig02CondFrac}(c) can also be alternatively understood from the pairing correlations. Fig.~\ref{fig:Fig09StruVsMu} plots the vertex contributions of the pairing structure factors $\bar{S}_{\ell}(\boldsymbol{\Gamma})$ versus fermion filling. The results of on-site singlet structure factor has the same nonmonotonic behavior as the singlet condensate fraction in Fig.~\ref{fig:Fig02CondFrac}(c). Instead, the NN and NNN triplet structure factors show more interesting signatures with different peak locations, revealing that the triplet channel is first governed by the NNN and then by the NN pairing from half filling to low filling regime ($n<0.5$). These validate the results of condensate fractions in Fig.~\ref{fig:Fig02CondFrac}(c) and pair wave functions in Fig.~\ref{fig:Fig03PairWfVsMu}. Moreover, the wide plateau of the triplet condensate fraction in Fig.~\ref{fig:Fig02CondFrac}(c) can be explained by the accumulated results of the NN and NNN triplet pairing correlations in Fig.~\ref{fig:Fig09StruVsMu}(b). 

As for the temperature and interaction strength, we have also obtained the vertex contributions of both singlet and triplet pairing correlations. In Appendix~\ref{sec:AppendixB}, we have presented the results of vertex contributions $\bar{P}_{\ell}(\mathbf{r})$ versus temperature. 

\subsection{BKT transition temperature from condensate fractions}
\label{sec:ComputeBKT}

In previous studies of 2D attractive Hubbard model, the BKT transition temperature was usually determined by the finite-size scaling of the pairing structure factor or from the universal jump property of the superfluid density~\cite{Paiva2004,Fontenele2022,Tang2014}. However, these quantities can become significantly small for low filling system, which makes the finite-size scaling even harder. In Ref.~\cite{Tang2014}, the BKT transition temperatures for the same system as we study were calculated from superfluid density with systems up to $L=12$. Such AFQMC simulations need to compute the superfluid density from dynamical current-current correlation functions, which definitely cost much more computational effort to reach high-precision results. 

Alternatively, numerical studies in 2D XY models solidly confirm that the finite-size BKT transition temperature $T_{\rm BKT}(L)$ has a following form~\cite{Tomita2002,Nguyen2019} as
\begin{equation}\begin{aligned}
\label{eq:BKTtempL}
T_{\rm BKT}(L) = T_{\rm BKT}(L=\infty) + \frac{a}{(\ln L + b)^2},
\end{aligned}\end{equation}
with $a,b$ as coefficients related to the specific problem, and $T_{\rm BKT}(L=\infty)$ as the final answer under thermodynamic limit. The second term in Eq.~(\ref{eq:BKTtempL}) containing the logarithm of linear system size $L$ already indicates the strong finite-size effect. As a result, biased result of $T_{\rm BKT}(L=\infty)$ might be obtained if only a small group of systems with not large enough sizes are accessed in the calculations. Based on Eq.~(\ref{eq:BKTtempL}), we could extrapolate the precise $T_{\rm BKT}(L=\infty)$ from the finite-size $T_{\rm BKT}(L)$ results. In the previous study of the 2D interacting Fermi gas without SOC~\cite{Yuanyao2022}, it was found that the first-order derivative of condensate fraction over the temperature shows a peak and its location can be identified as $T_{\rm BKT}(L)$. Such calculations do not involve dynamical measurements and are thus computationally much cheaper and high-precision results of $T_{\rm BKT}(L)$ can be yielded. Similar formula as Eq.~(\ref{eq:BKTtempL}) was also confirmed dealing with the convergence of $T_{\rm BKT}$ with number of fermions for 2D Fermi gas in Ref.~\cite{Yuanyao2022}. Thus, in the following, we also concentrate on calculating $T_{\rm BKT}(L)$ from the condensate fractions and reaching $T_{\rm BKT}(L=\infty)$ using Eq.~(\ref{eq:BKTtempL}), for the system with SOC.

\begin{figure}[ht]
\centering
\includegraphics[width=1.0\columnwidth]{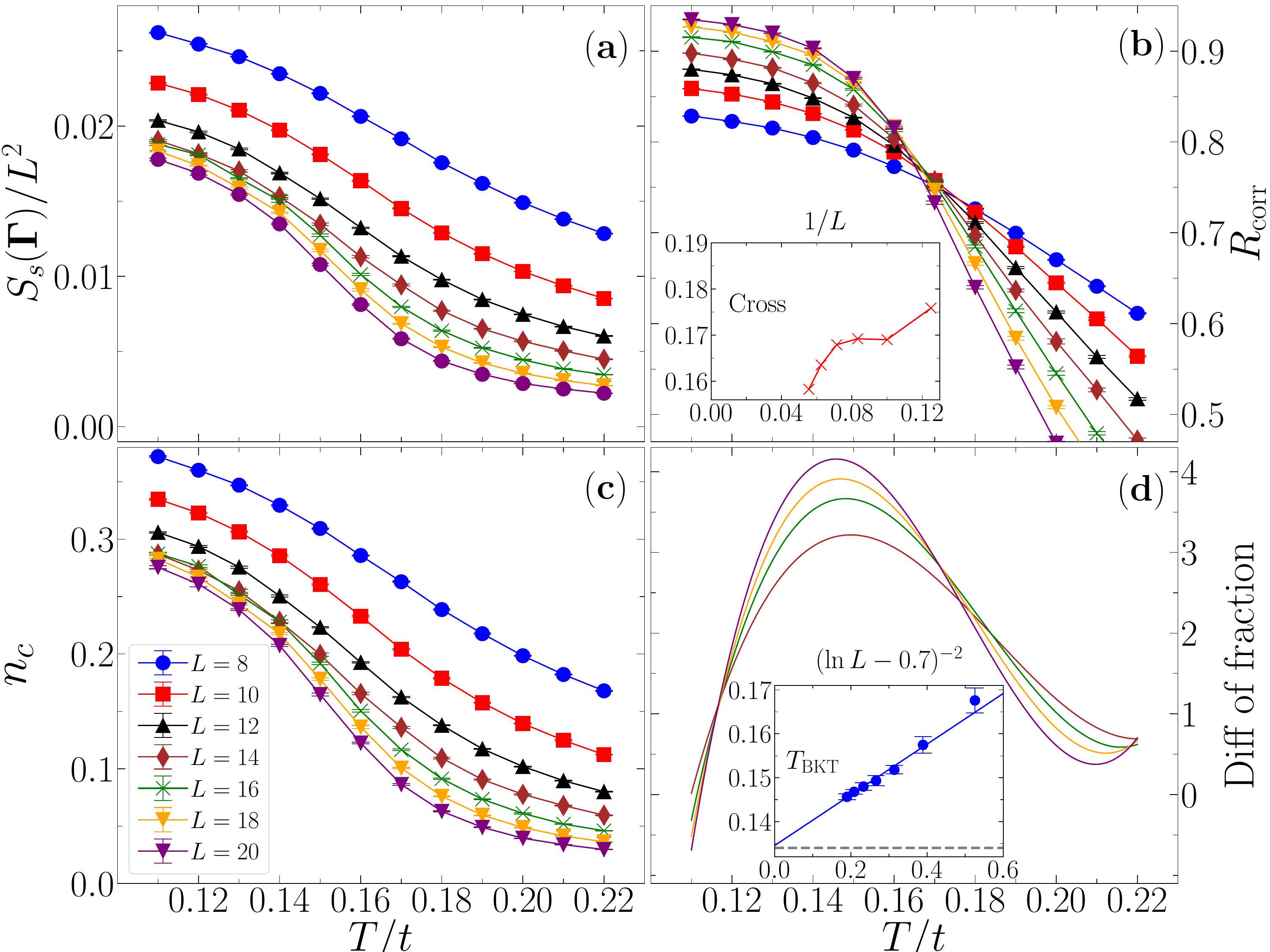}
\caption{\label{fig:Fig10BKT01} Determination of the BKT transition temperatures from correlation ratio and total condensate fraction. (a)(b) are the singlet pairing structure factor $S_s(\boldsymbol{\Gamma})$ and the corresponding correlation ratio. The inset in panel (b) plots the cross points of finite-size correlation ratios. (c)(d) are the total condensate fraction and its first order derivative (after polynomial fitting). The inset in panel (d) plots $T_{\rm BKT}(L)$ after the best fitting using Eq.~(\ref{eq:BKTtempL}), reaching the final result as $T_{\rm BKT}(L=\infty)/t=0.135(4)$. These calculations are performed for $L=8\sim 20$ systems with $U/t=-4,\lambda/t=0.5,\mu/t=0.5$.}
\end{figure}

In Fig.~\ref{fig:Fig10BKT01}, we first demonstrate the determination of BKT transition temperature from the spin-singlet pairing structure factor $S_s(\boldsymbol{\Gamma})$ (defined in Sec.~\ref{sec:TripletCrFt}) and the total condensate fraction as a comparison. The correlation ratio for $S_s(\boldsymbol{\Gamma})$ is defined as $R_{\rm corr}=1 - S_s(\boldsymbol{\Gamma}+\mathbf{q})/S_s(\boldsymbol{\Gamma})$ with $\mathbf{q}$ as the smallest momentum on the lattice, i.e., $(2\pi/L,0)$ or $(0,2\pi/L)$. It resembles the Binder cumulant which converges to unity in ordered phase and vanishes in the disordered phase in thermodynamic limit. Then the cross points of the finite-size $R_{\rm corr}$ results can be approximately viewed as the transition temperature. As shown in Fig.~\ref{fig:Fig10BKT01}(b), the cross points of $R_{\rm corr}$ indeed move to the lower temperature with system size but not with a well defined behavior. Instead, for the total condensate fraction in Fig.~\ref{fig:Fig10BKT01}(c), we first perform a polynomial fitting to the numerical data and then compute its first-order derivative and get the location of the peak as $T_{\rm BKT}(L)$ [shown in Fig.~\ref{fig:Fig10BKT01}(d)], which avoids the step error involved in the numerical derivative. We have further calculated the error bars of $T_{\rm BKT}(L)$ applying the standard bootstrapping technique. Finally, we use Eq.~(\ref{eq:BKTtempL}) to extrapolate the final result of BKT transition temperature $T_{\rm BKT}(L=\infty)=0.135(4)$, as plotted in inset of Fig.~\ref{fig:Fig10BKT01}(d). The details of the bootstrapping calculations of $T_{\rm BKT}(L)$ are presented in Appendix~\ref{sec:AppendixC}. These results also indicates large finite-size effect in $R_{\rm corr}$ as the cross point of $L=18$ and $L=20$ is $T/t=0.158$, which strongly deviates from $T_{\rm BKT}(L=\infty)$.

\begin{figure}[ht]
\centering
\includegraphics[width=1.0\columnwidth]{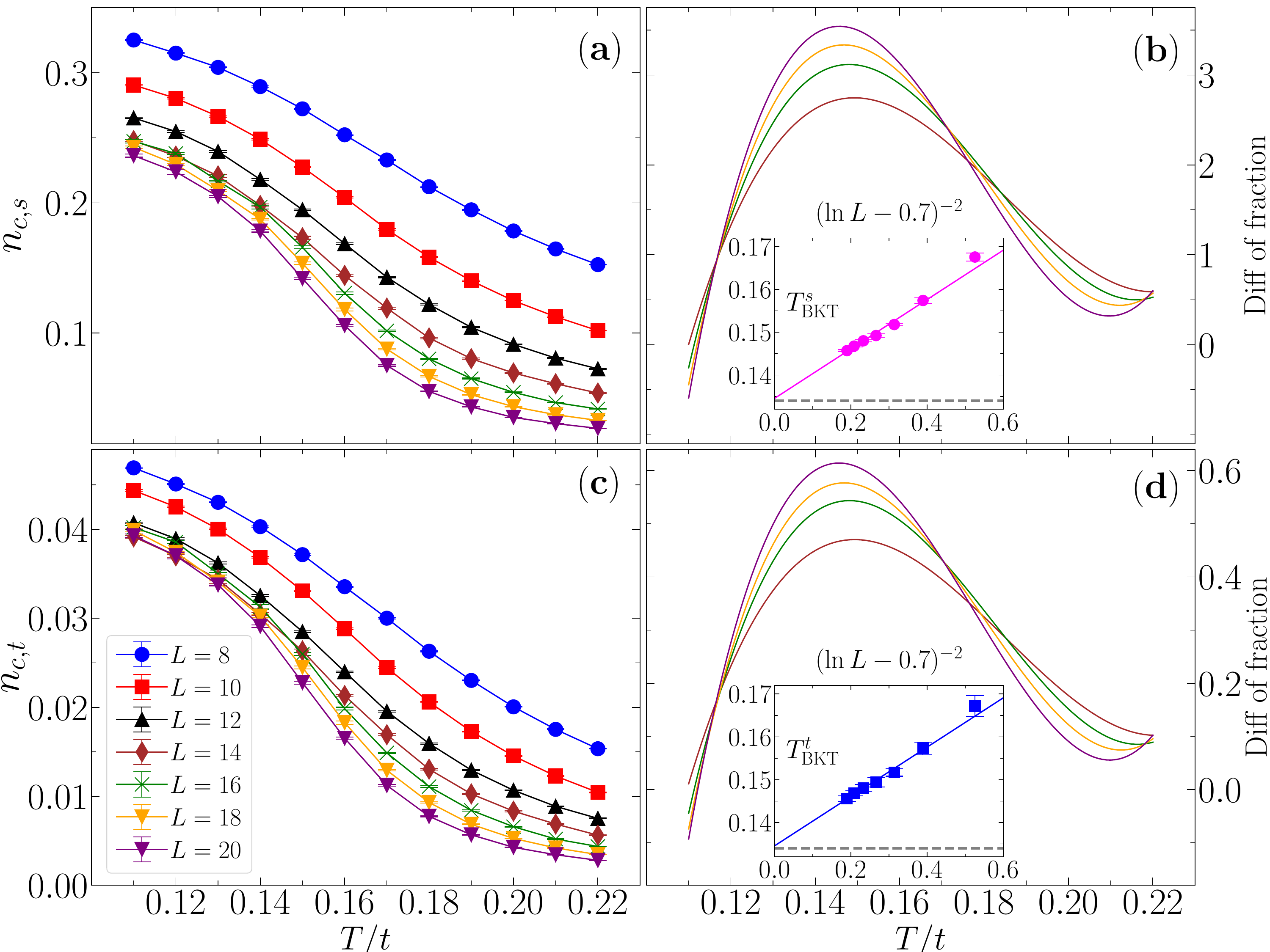}
\caption{\label{fig:Fig11BKT02} Determination of the BKT transition temperatures from singlet and triplet condensate fractions. (a)(b) are the singlet condensate fraction and its first-order derivative (after polynomial fitting). (c)(d) are the triplet condensate fraction and its first-order derivative. The inset in panel (b)(d) plots $T_{\rm BKT}(L)$ after the best fitting using Eq.~(\ref{eq:BKTtempL}), reaching the final result as $T^{s}_{\rm BKT}(L=\infty)/t=0.135(4)$ and $T^{t}_{\rm BKT}(L=\infty)/t=0.135(4)$, respectively. Simulation parameters are the same as Fig.~\ref{fig:Fig10BKT01}.}
\end{figure}

For the mixed-parity pairing we stuty, we also have the numerical data of singlet and triplet condensate fractions. From them, we can separately extrapolate the BKT transition temperatures for the spin-singlet and triplet superfluidity as $T_{\rm BKT}^s(L=\infty)$ and $T_{\rm BKT}^t(L=\infty)$, which are expected be the same. In Fig.~\ref{fig:Fig11BKT02}, we illustrate the determination of $T_{\rm BKT}$ from both singlet and triplet condensate fractions. The procedure is exactly the same as that in Fig.~\ref{fig:Fig10BKT01}(c) and (d), and the details can also be refered in Appendix~\ref{sec:AppendixC}. These calculations produce the final results of $T_{\rm BKT}^s(L=\infty)=0.135(4)$ and $T_{\rm BKT}^t(L=\infty)=0.135(4)$. These results are indeed consistent as expectation, meaning the BKT transition for the quasi-long-range mixed-parity pairing order happens simultaneously for singlet and triplet channels. 

Based on the results in Fig.~\ref{fig:Fig10BKT01} and Fig.~\ref{fig:Fig11BKT02}, we have obtained the BKT transition temperature $T_{\rm BKT}(L=\infty)=0.135(4)$ for the parameter $U/t=-4,\lambda/t=0.5,\mu/t=0.5$ (with fermion filling $n=0.6795$ at the transition point). This result is consistent with the $T_{\rm BKT}$ computed for filling $n=0.7$ in Ref.~\cite{Tang2014}. We can then conclude that, similar to previous studies~\cite{Yuanyao2022}, it's also an efficient way to determine BKT transition temperature from (total, singlet and triplet) condensate fractions for attractive fermion systems with SOC. 

\section{Summary and Discussion}
\label{sec:Summary}

The mixed-parity pairing phenomena is theoretically a natural result for fermionic systems with broken inversion symmetry~\cite{Bychkov1984,Rashba2001}, and it has been experimentally observed in various three-dimensional superconductors with SOC~\cite{Smidman2017}. In addition, the experimental realization of SOC with an artificial gauge field in optical lattice by ultracold atoms~\cite{Huang2016,Meng2016,Sun2018} provides the opportunity to perform more systematic and deeper study of the mixed-parity pairing in a more controlled manner. Our AFQMC numerical results in this work can not only serve as quantitative guide for such 2D optical lattice experiments, but also present some new physical results on the essential pairing structure of the corresponding mixed-parity pairing.

In summary, we have applied the numerically exact finite-temperature AFQMC method to study the pairing properties of attractive fermions with Rashba SOC in 2D optical lattice. We evaluate the contributions of the spin-singlet and triplet channels to the mixed-parity pairing. With the scanning of temperature, fermion filling, SOC and interaction strengths, we find that the singlet pairing plays a dominant role with relatively small triplet contribution in most relevant parameter regimes. From the pair wave functions, we find that, for intermediate interaction ($U/t=-4$), the singlet pairing mainly consists of local Cooper pairs while the triplet channel is rather extended with major contributions from several nearest neighbors. Especially, in low filling regime ($n<0.5$), the triplet pairing is dominated by NN fermion pairs, in contrast with the NNN ones around half filling. Via the vertex contribution of pairing correlations, we have shown that the triplet pairing is first enhanced and then suppressed with increasing SOC, and there exists an optimal SOC strength for observing the triplet pairing. Finally, we have demonstrated the computation of the BKT transition temperature from the finite-size results of total, singlet and triplet condensate fractions, suggesting it also as an efficient method for systems with SOC. Our numerical results will surely offer useful benchmarks for future optical lattice experiments as well as theories and other numerical methods.

Our work also has implications for achieving the spin-triplet superconductivity and superfluidity. Considering the fact that the triplet pairing is only confirmed to exist in very rare systems, it might be a way out to pay more attention to the systems with mixed-parity pairing. Specifically, if one can control the triplet contribution to the pairing by tunning physical parameters (for example, the SOC strength) in such systems, we might access the special case in which the triplet channel dominates, similar to $\text{Li}_2\text{Pt}_3\text{B}$~\cite{Nishiyama2007}. Unfortunately, our work shows it's very unlikely to realize such special case for the system described by the lattice model in Eq.~(\ref{eq:2DHamlt}). Instead, there are actually other possibilities, such as further including the Dresselhaus SOC, and NN or NNN attractive interactions. The former was found to be useful in promotion of the triplet contribution in interacting Fermi gas within the mean-field theory~\cite{Anna2011,*Anna2012}. The latter is apparently supported by our numerical results as the triplet pairing is mainly contributed by NN and NNN Cooper pairs. We leave these open possibilities to future work.

\begin{acknowledgments}
Y.Y.H. acknowledges Peter Rosenberg and Shiwei Zhang for valuable discussions. This work was supported by the National Natural Science Foundation of China (under Grant No. 12047502, 12204377 and 12275263), the Innovation Program for Quantum Science and Technology (under Grant No. 2021ZD0301900), and the Youth Innovation Team of Shaanxi Universities.
\end{acknowledgments}

\appendix
\section{Structure factor of the density-density correlation function}
\label{sec:AppendixA}

In Ref.~\cite{Rosenberg2017}, it was found that, at half filling, the long-range charge density wave (CDW) order with checkerboard pattern coexist with the pairing order in ground state for the lattice model in Eq.~(\ref{eq:2DHamlt}). We have also checked this and our numerical results suggest that the long-range CDW order should not exist for the case away from half filling. 

We compute the density-density correlation function defined as $D(\mathbf{r})=\frac{1}{4}(\langle\hat{n}_{\mathbf{i}}\hat{n}_{\mathbf{i}+\mathbf{r}}\rangle - \langle\hat{n}_{\mathbf{i}}\rangle\langle\hat{n}_{\mathbf{i}+\mathbf{r}}\rangle)$ (with $\hat{n}_{\mathbf{i}}=\hat{n}_{\mathbf{i}\uparrow}+\hat{n}_{\mathbf{i}\downarrow}$), and the corresponding momentum-space structure factor as $S_{\rm CDW}(\mathbf{q})=\sum_{\mathbf{r}}D(\mathbf{r})e^{i\mathbf{q}\cdot\mathbf{r}}$. The leading component of $S_{\rm CDW}(\mathbf{q})$ appears at $\mathbf{q}=\mathbf{M}=(\pi,\pi)$ point, consistent with the CDW order with the checkerboard pattern.

\begin{figure}[ht]
\centering
\includegraphics[width=1.0\columnwidth]{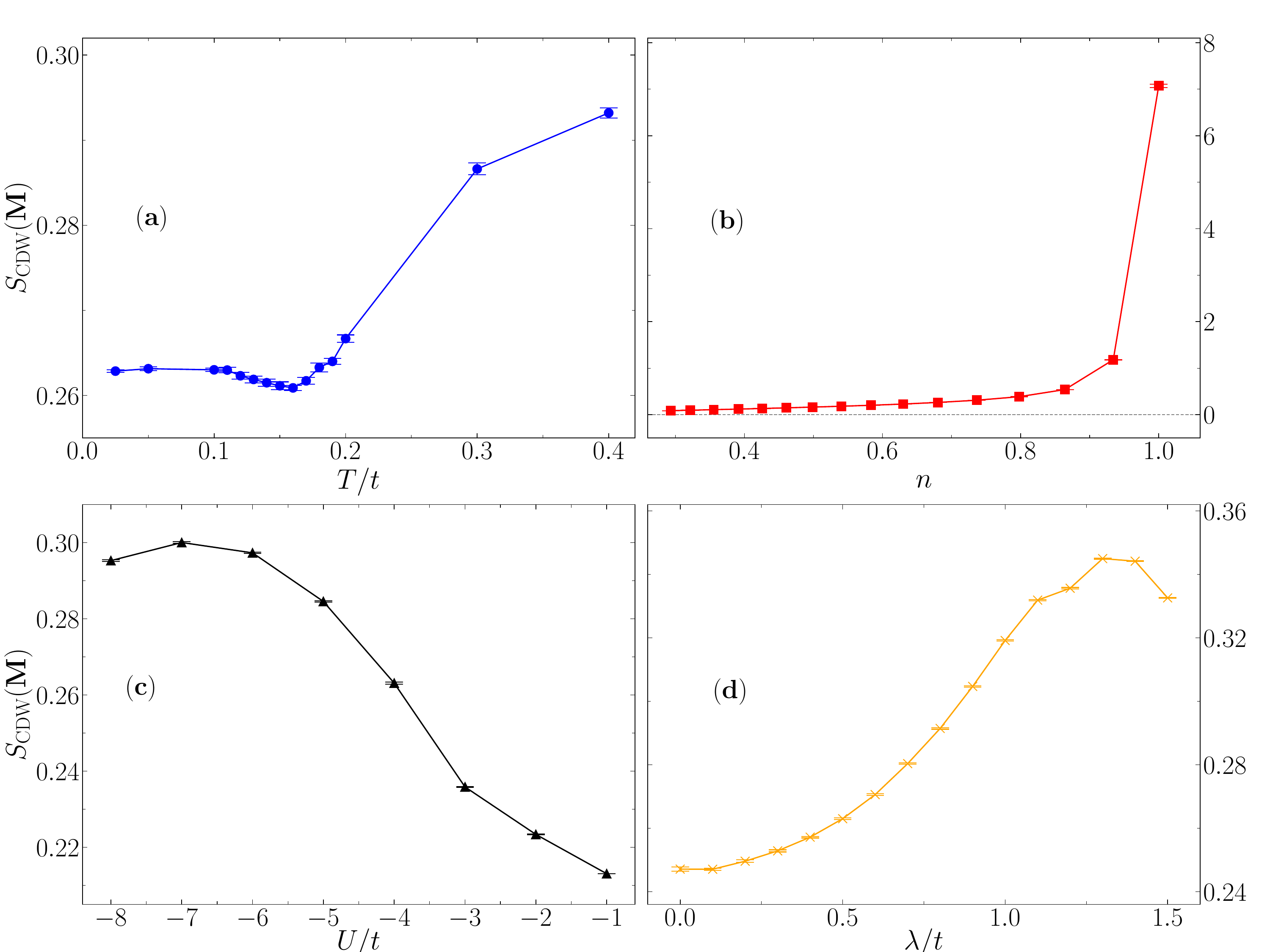}
\caption{\label{fig:FigS01CDWOrder} The density-density structure factor at $\Gamma$ point for different quantities. The detailed parameters are shown in the figure. Compared with half-filling case ($\mu=0$), the structure factor has an obvious decrease, which indicates that the superconducting order and CDW do not coexist in the doped system. }
\end{figure}

In Fig.~\ref{fig:FigS01CDWOrder}, we illustrate the results of the CDW structure factor $S_{\rm CDW}(\mathbf{M})$ with various tunning parameters. First, with doping as increasing the chemical potential, $S_{\rm CDW}(\mathbf{M})$ immediately decrease from the half-filling result by approximately an order of magnitude for $n=0.94$, which suggests the significant suppression of CDW order away from half filling. Second, the results with lowering temperature with $\mu/t=0.5$ (around $n=0.68$) explicitly shows that $S_{\rm CDW}(\mathbf{M})$ first decreases, then reaches a minimum and gradually saturates towards $T=0$, indicating the absence of long-range order. The results with varying SOC and interaction strengths show some enhancements of $S_{\rm CDW}(\mathbf{M})$ for specific regimes, but its largest values are still much smaller than the half-filling results, which also suggests the short-range correlations.

\section{Vertex contributions of the pairing correlation functions versus temperature}
\label{sec:AppendixB}

In Sec.~\ref{sec:TripletCrFt}, we have shown the numerical results of vertex contributions for the pairing correlation functions with tunning SOC strength and chemical potential. Here, we present more results with varying temperature. 

\begin{figure}[ht]
\centering
\includegraphics[width=1.0\columnwidth]{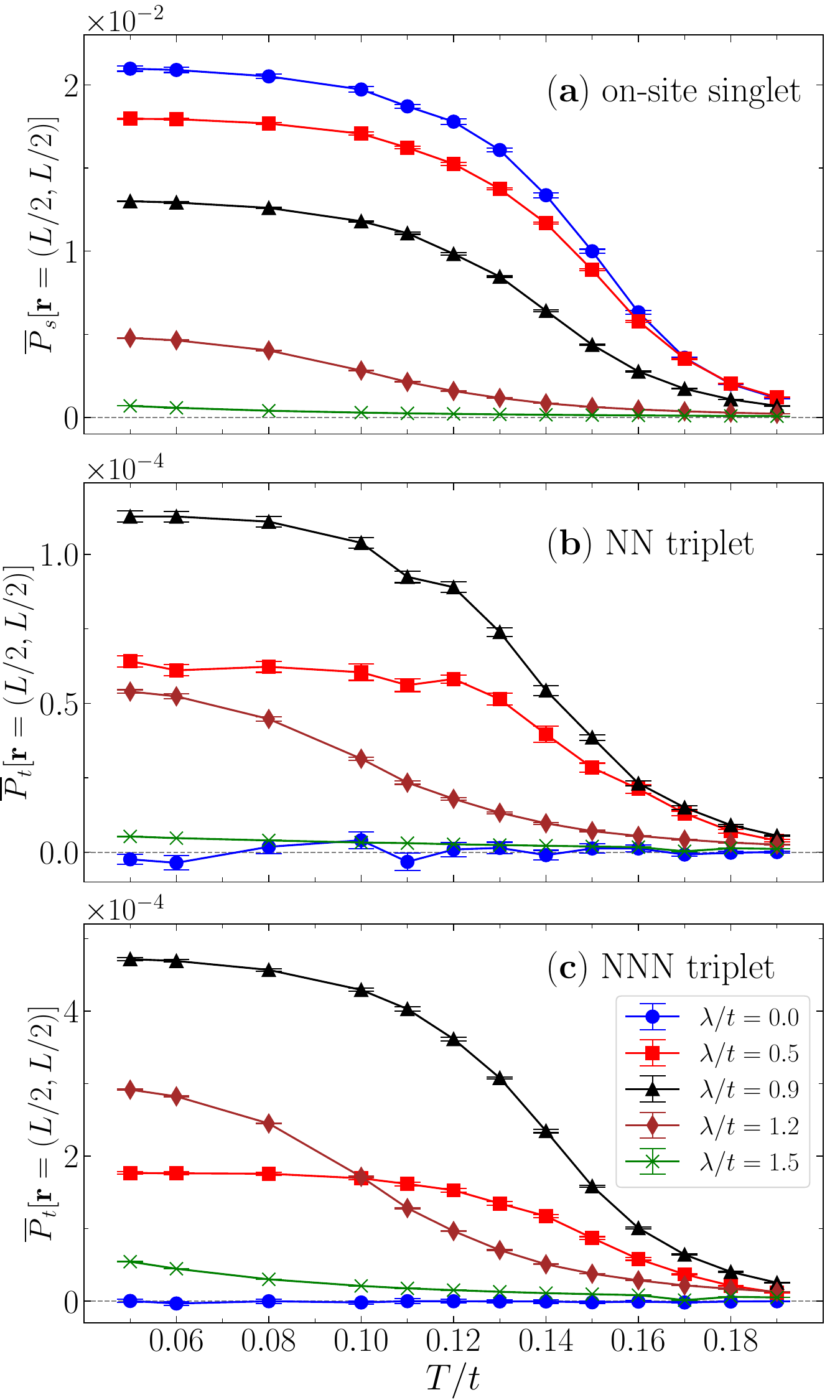}
\caption{\label{fig:FigS02CrFVsTemp} Vertex contribution of on-site singlet (top), NN triplet (middle) and NNN triplet (bottom) pairing correlation functions $P_{\ell}(\mathbf{r})$ (with $\ell=s,t$) versus temperature, with $r=\sqrt{2}L/2=10\sqrt{2}$ [as $\mathbf{r}=(L/2, L/2)$] as the largest distance. Results with several SOC strengths $\lambda/t=0\sim 1.5$ are plotted. These calculations are performed for $L=20$ system with $T/t=0.10$ and $U/t=-4$, $\mu/t=0.5$.}
\end{figure}

In Fig.~\ref{fig:FigS02CrFVsTemp}, we present the vertex of reals-space pairing correlations $P_{\ell}(\mathbf{r})$ (with $\ell=s,t$) with the largest distance as $r=\sqrt{2}L/2$ [as $\mathbf{r}=(L/2, L/2)$] on the lattice with on-site singlet, NN and NNN triplet channels versus the temperature, for several SOC strengths as $\lambda/t=0\sim 1.5$. All the results show enhancements with decreasing temperature and plateaus appear saturating to the $T=0$ results, indicating the quasi-long-range pairing order at low temperature regime. At low temperature regime, the triplet correlations reach the maximum at $\lambda/t=0.9$ for both NN and NNN pairing, consistent with the results shown in Fig.~\ref{fig:Fig07CrFVsSOC}. Besides, these results also illustrate that the results at $T/t=0.10$ is very close to the $T=0$ correspondences.

\section{The determination of the BKT transition temperature \texorpdfstring{$T_{\rm BKT}(L)$}{text} and \texorpdfstring{$T_{\rm BKT}(L=\infty)$}{text}}
\label{sec:AppendixC}

In this section, we present the details for the determination of the BKT transition temperature $T_{\rm BKT}(L)$ and $T_{\rm BKT}(L=\infty)$. 

\begin{figure}[ht]
\centering
\includegraphics[width=1.0\columnwidth]{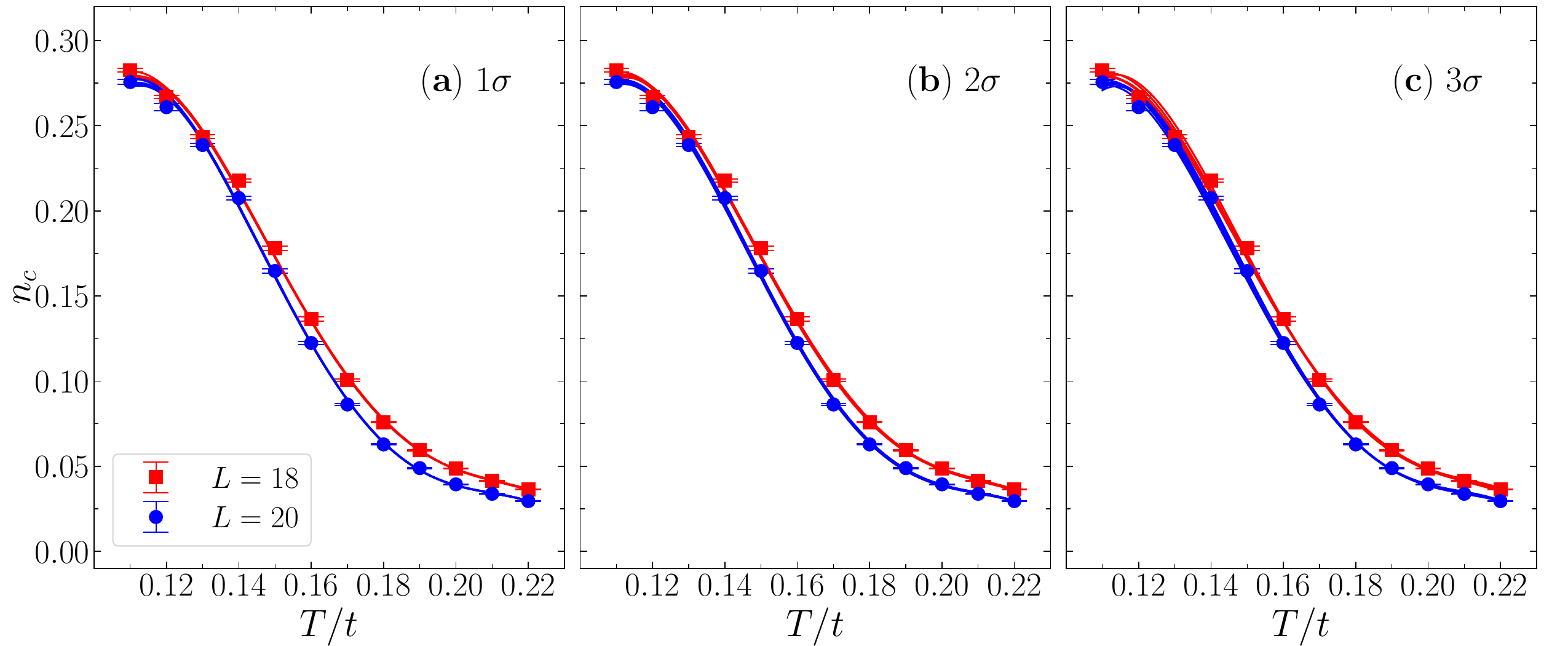}
\caption{Condensate fraction versus temperature. The points are QMC data for system size $L=18$ and $20$, where the error bars denote the standard error. The curves are the fitting results of different random $n_{c,i}(L,T,q) = \bar{n}_{c}(L,T) + N(0,q\sigma(L,T))$. For simplicity, we plot five random curves for each system. The parameters are $U/t=-4$, $\lambda/t=0.5$ and $\mu/t=0.5$.}
\label{fig:func}
\end{figure}

Based on the numerical data of condensate fraction $\bar{n}_{c}(L,T)$ (including the total, singlet and triplet) and the corresponding standard error $\sigma(L,T)$, we apply the bootstrapping technique by first generating a set of random data by
\begin{equation}\begin{aligned}
n_{c,i}(L,T,q) = \bar{n}_{c}(L,T) + N(0,q\sigma(L,T)),
\end{aligned}\end{equation}
where $i$ denotes the $i$-th random data with $q=1,2,3$ for different range of deviation, and $N(0,q\sigma(L,T))$ stands for the normalized Gaussian distribution with expectation and standard deviation as $0$ and $q\sigma(L,T)$. The whole process follows the Gauss analysis and can quickly construct a large number of $n_{c,i}(L,T,q)$. Then we fit $n_{c,i}(L,T,q)$ for every set of random data with a fourth-order polynomial of temperature around the transition point, and compute the peak location of its first-order derivative, and then we take it as $T_{\rm BKT}(i,L,q)$. With the full set of $T_{\rm BKT}(i,L,q)$, one can perform data analysis and obtain $\overline{T}_{\rm BKT}(L,q)$ with the stand deviation as its error bar. Compared with the method to only fit $n_c(L,T)$ with the original data, this bootstrapping method can additional present a reasonable error bar for $\overline{T}_{\rm BKT}(L,q)$. 

\begin{figure}[t]
\centering
\includegraphics[width=0.99\columnwidth]{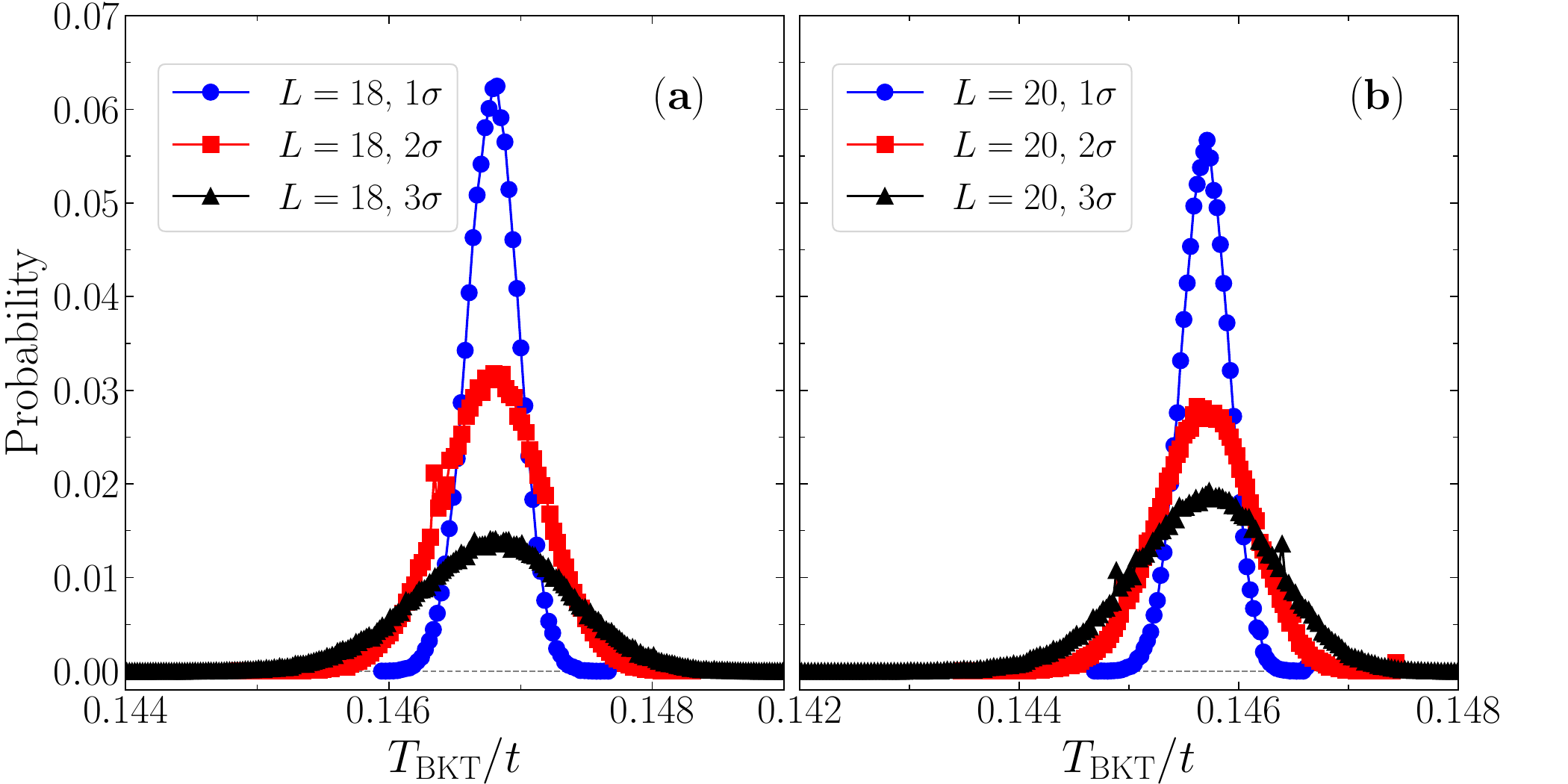}
\caption{The distribution of $T_{\rm BKT}(L,q)$ based on the bootstrapping calculations. We have generated 150000 random data for each system size and $q$. It is well illustrated that the average values $\overline{T}_{\rm BKT} (L)$ for different $q$ are identical. Simulation parameters are the same as Fig.~\ref{fig:func}.}
\label{fig:2fig}
\end{figure}

\begin{figure}[ht]
\centering
\includegraphics[width=1.0\columnwidth]{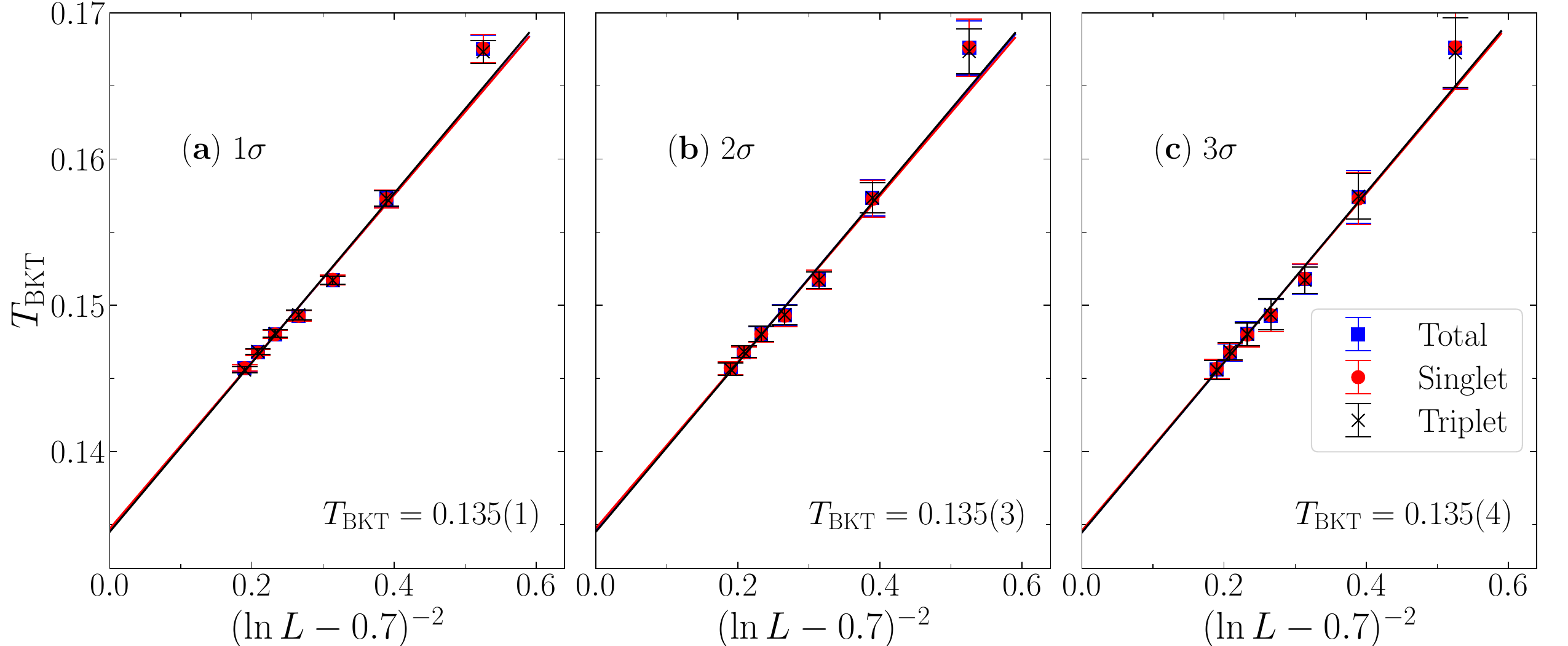}
\caption{BKT transition temperature $T_{\rm BKT}(L,q)$ versus $(\ln L-0.7)^{-2}$ for the total, singlet and triplet condensate fractions. The fittings are based on the correction formula Eq.~(\ref{eq:BKTtempL}), where $-0.7$ is determined from the fitting. 
The singlet and triplet channels give very similar results. Simulation parameters are the same as Fig.~\ref{fig:func}.} 
\label{fig:lss}
\end{figure}

In order to show the process, we take $L=18$ and $20$ for example. Fig.~\ref{fig:func} shows the original data and fitting polynomials of five random sets of data. It is shown that the fourth-order polynomials can capture the essential behavior of $n_c$ around the transition point. By generating 150000 samples, Fig.~\ref{fig:2fig} shows the histogram of results for $\overline{T}_{\rm BKT} (L,q)$, which are fairly consistent with Gaussian distributions. It is also well illustrated that, for different $q$, the average $\overline{T}_{\rm BKT}(L,q)$ are obviously identical for both system sizes. The only difference of these results the data set generated by different Gaussian noise (different $q$) shows in the standard deviations of these distributions. As espectation, the distribution is wider (with larger standard deviation) for larger $q$.

Finally, to obtain $T_{\rm BKT}$ in the thermodynamic limit, we perform fittings of $T_{\rm BKT} (L,q)$ with the formula in Eq.~(\ref{eq:BKTtempL}). Fig.~\ref{fig:lss} shows the fitting results with different $q$. The total condensate fraction as well as its two channels give similar $T_{\rm BKT}$, as shown in Fig.~\ref{fig:lss}. To achieve a confident estimate of $T_{\rm BKT}(L=\infty)$, we adopt the results of $q=3$ as the final result as presented in Sec.~\ref{sec:ComputeBKT} of the main text.

\bibliography{SOCPairingMain}

\end{document}